\documentclass[preprint,superscriptaddress,12pt]{revtex4}
\usepackage{graphicx}
\usepackage{booktabs}

\usepackage[normalem]{ulem}

\begin{document}
\title{Detection of Crashes and Rebounds in Major Equity Markets}

\author{Wanfeng Yan}
 \email{wyan@ethz.ch}
 \affiliation{Department of Management, Technology and Economics, ETH Zurich, Kreuzplatz 5, CH-8032 Zurich, Switzerland} %

\author{Reda Rebib}
 \email{reda.rebib@gmail.com}
 \affiliation{Department of Management, Technology and Economics, ETH Zurich, Kreuzplatz 5, CH-8032 Zurich, Switzerland} %

\author{Ryan Woodard}
 \email{rwoodard@ethz.ch}
 \affiliation{Department of Management, Technology and Economics, ETH Zurich, Kreuzplatz 5, CH-8032 Zurich, Switzerland} %

\author{Didier Sornette}
 \email{dsornette@ethz.ch}
 \affiliation{Department of Management, Technology and Economics, ETH Zurich, Kreuzplatz 5, CH-8032 Zurich, Switzerland} %
 \affiliation{Swiss Finance Institute, c/o University of Geneva, 40 blvd. Du Pont d'Arve, CH-1211 Geneva 4, Switzerland}%

\begin{abstract}
  Financial markets are well known for their dramatic dynamics and consequences
  that affect much of the world's population.  Consequently, much research has
  aimed at understanding, identifying and forecasting crashes and rebounds in
  financial markets. The Johansen-Ledoit-Sornette (JLS) model provides an
  operational framework to
  understand and diagnose financial bubbles from rational expectations and was recently extended
  to negative bubbles and rebounds. Using the  JLS model, we
  develop an alarm index based on an advanced pattern recognition method
  with the aim of detecting bubbles and performing forecasts of
  market crashes and rebounds. Testing our methodology on
  10 major global equity markets, we show quantitatively that our developed alarm performs much better than chance
  in forecasting market crashes and rebounds. We use the derived signal to develop
  elementary trading strategies that produce statistically better performances than a
  simple buy and hold strategy.

{\bf Keywords:} the JLS model, financial bubbles, crashes, rebounds,
log-periodic power law, pattern recognition method, alarm index,
prediction, error diagram, trading strategy.

\end{abstract}
\keywords{}

 \maketitle \clearpage

\section{Introduction} \label{sec:introduction}

Hundreds of millions of people are influenced by bubbles and crashes in the
financial system. Although many academics, practitioners and policy makers have
studied these extreme events, there is little consensus yet about their causes
or even definitions (see Ref.\cite{kaizoji-bubble-08} for a review of
existing models and references therein). More than a decade ago,
the Johansen-Ledoit-Sornette (JLS)
model~\cite{jls,jsl,js,sornettecrash} has been developed to detect bubbles and
crashes in financial markets. This model states that
imitation and herding behavior of noise traders during the bubble regime may
lead to transient accelerating price growth that may end with a possible crash. Before the
crash, the price grows at a faster-than-exponential rate rather than an
exponential rate, which is normally used in other bubble
models~\cite{Watanabe-Takayasu-07,abreubrunnermeier}.  This
faster-than-exponential growth is due to positive feedback in the valuation of
assets among the investors. Further consideration of tension and competition
between the fundamentalists and the noise traders leads the real prices to
oscillate about this growth rate with increasing frequency.  This oscillation
is actually periodic in the logarithm of time before the most likely end of the
bubble growth and, therefore, the model describes log-periodic power law (LPPL)
behavior.

Since positive feedback mechanisms in the markets may also lead to
transient periods of accelerating \emph{decrease} in price, followed in turn by
rapid reversals, the JLS model was recently extended to study ``negative
bubbles'' and rebounds~\cite{rebound,phpro}.  There, it was shown that, given the
existence of extended crashes and rebounds, negative bubbles could be
identified and their endings forecast in the same spirit as for ``traditional''
bubbles such as the 2006-2008 oil bubble \cite{oil}, the Chinese index bubble
in 2009 \cite{Jiangetal09}, the real estate market in Las Vegas
\cite{ZhouSorrealest08}, the U.K. and U.S. real estate bubbles
\cite{Zhou-Sornette-2003a-PA,Zhou-Sornette-2006b-PA}, the Nikkei index
anti-bubble in 1990-1998 \cite{JohansenSorJapan99}, the S\&P 500 index
anti-bubble in 2000-2003 \cite{SorZhoudeeper02}, the Dow Jones Industrial
Average historical bubbles
\cite{Vandewalle-Ausloos-Boveroux-Minguet-1999-EPJB}, the corporate bond
spreads \cite{Clark-2004-PA}, the Polish stock market bubble
\cite{Gnacinski-Makowiec-2004-PA}, the western stock markets
\cite{Bartolozzi-Drozdz-Leinweber-Speth-Thomas-2005-IJMPC}, the Brazilian real
(R\$) - US dollar (USD) exchange rate
\cite{Matsushita-daSilva-Figueiredo-Gleria-2006-PA}, the 2000-2010 world major
stock indices \cite{Drozdz-Kwapien-Oswiecimka-Speth-2008-APPA}, the South
African stock market bubble \cite{ZhouSorrSA09} and the US repurchase
agreements market \cite{repo}. Moreover, new experiments in ex-ante bubble
detection and forecast were recently launched in the Financial Crisis
Observatory at ETH Zurich \cite{BFE-FCO09, BFE-FCO10, BFE-FCO11}.

To test this idea and to implement a systematic forecast procedure based on the
JLS model, Sornette and Zhou adapted a pattern recognition method to detect
bubbles and crashes~\cite{SorZhouforecast06}. This method was originally
developed by mathematician I. M. Gelfand and his collaborators in the mid-1970's
as an earthquake prediction scheme.
Since then, this method has been widely used in many kinds of predictions,
ranging from uranium prospecting \cite{briggs_pattern_1977} to unemployment
rates \cite{keilis-borok_patterns_2005}. Yan, Woodard and Sornette
\cite{rebound,phpro} extended this method to negative bubbles and rebounds of
financial markets. They also improved the method in~\cite{SorZhouforecast06} by
separating the learning period and prediction period to enable a pure causal
prediction.

Since the study~\cite{rebound,phpro} only tested for rebounds in one major
index (S\&P 500), in this paper, we expand to ten major global equity markets
using the pattern recognition method to detect and forecast crashes and
rebounds. Our results indicate that the performance of the predictions on both
crashes and rebounds for most of the indices is better than chance.  That is,
the end of large drawups and drawdowns and the subsequent crashes and rebounds can be successfully
forecast. To demonstrate this, we design a simple trading strategy and show
that it out-performs a simple buy-and-hold benchmark.

The structure of the paper is as follows. In Section~\ref{sec:jls}, we briefly
introduce the JLS model and the bubble/negative bubble versions of the model.
Then, we present the pattern recognition method for the prediction of crashes
and rebounds in Section~\ref{sec:pr}. The quality of the prediction is tested
in Section~\ref{sec:prediction} using error diagrams to compare missed events
versus total alarm time. We next introduce the trading strategy based on the
alarm index and test its performance in Section~\ref{sec:strategy}. We
summarize our results and conclude in Section~\ref{sec:conclusion}.

\section{The JLS Model for Detecting Bubbles and Negative
  Bubbles} \label{sec:jls}

The JLS model \cite{js,jsl,jls,sornettecrash} is an extension of the rational
expectation bubble model of Blanchard and Watson \cite{bwrationalexpectation}. In this model, a
financial bubble (negative bubble) is modeled as a regime of super-exponential
power law growth (decline) punctuated by short-lived corrections organized
according to the symmetry of discrete scale invariance \cite{DSI-sornette98}.
The super-exponential power law is argued to result from positive feedback
resulting from noise trader decisions that tend to enhance deviations from
fundamental valuation in an accelerating spiral.  That is, the price of a stock
goes higher (lower) than the fundamental value during the bubble (negative
bubble) period, ending in a sudden regime change.

In the JLS model, the dynamics of asset prices is described as
\begin{equation}
  \frac{dp}{p} = \mu(t)dt + \sigma(t)dW - \kappa dj~,
  \label{eq:dynamic}
\end{equation}
where $p$ is the stock market price, $\mu$ is the drift (or trend) and $dW$ is
the increment of a Wiener process (with zero mean and unit variance). The term
$dj$ represents a discontinuous jump such that $dj = 0$ before the crash or
rebound and $dj = 1$ after the crash or rebound occurs. The change amplitude
associated with the occurrence of a jump is determined by
$\kappa$. The parameter $\kappa$ is positive for bubbles and negative for
negative bubbles. The assumption of a constant jump size is easily relaxed by
considering a distribution of jump sizes, with the condition that its first
moment exists. Then, the no-arbitrage condition is expressed similarly with
$\kappa$ replaced by the mean of the distribution.  Each successive crash
corresponds to a jump of $dj$ by one unit. The dynamics of the jumps is
governed by a hazard rate $h(t)$. For the bubble (negative bubble) regime,
$h(t) dt$ is the probability that the crash (rebound) occurs between $t$ and
$t+dt$ conditional on the fact that it has not yet happened. We have $ E_t[dj]
= 1 \times h(t) dt + 0 \times (1- h(t) dt)$ and therefore
\begin{equation}
  {\rm E}_t[dj] = h(t)dt~.
  \label{theyjytuj}
\end{equation}
Under the assumption of the JLS model, noise traders exhibit collective herding
behaviors that may destabilize the market. The JLS model assumes that the
aggregate effect of noise traders should have a discrete scale invariant
property \cite{DSI-sornette98,sornettecrash}. Therefore, it can be accounted for by the
following dynamics of the hazard rate
\begin{equation}
  h(t) = B'(t_c-t)^{m-1}+C'(t_c-t)^{m-1}\cos(\omega\ln (t_c-t) -\phi')~.
  \label{eq:hazard}
\end{equation}

The no-arbitrage condition reads ${\rm E}_t[dp]=0$, where the expectation is
performed with respect to the risk-neutral measure and in the frame of the
risk-free rate. This is the standard condition that the price process is a
martingale. Taking the expectation of expression (\ref{eq:dynamic}) under the
filtration (or history) until time $t$ reads
\begin{equation}
{\rm E}_t[dp]= \mu(t) p(t) dt + \sigma(t) p(t) {\rm E}_t[dW] -
\kappa p(t) {\rm E}_t[dj]~.
 \label{thetyjye}
\end{equation}
Since ${\rm E}_t[dW] =0$ and  ${\rm E}_t[dj] = h(t)dt$ (equation
(\ref{theyjytuj})), together with the no-arbitrage condition ${\rm
E}_t[dp]=0$, this yields
\begin{equation}
\mu(t) = \kappa h(t)~. \label{tjyj4n}
\end{equation}
This result (\ref{tjyj4n}) expresses that the return $\mu(t)$ is
controlled by the risk of the crash quantified by its hazard rate
$h(t)$.

Now, conditioned on the fact that no crash or rebound has occurred, equation
(\ref{eq:dynamic}) is simply
\begin{equation}
  \frac{dp}{p} = \mu(t)dt + \sigma(t)dW = \kappa h(t) dt + \sigma(t)dW~.
  \label{eq:dynaiuhomic}
\end{equation}
Its conditional expectation leads to
\begin{equation}
{\rm E}_t\left[\frac{dp}{p}\right] =  \kappa h(t) dt~.
\end{equation}
Substituting with the expression (\ref{eq:hazard}) for $h(t)$ and
integrating yields the so-called log-periodic power law (LPPL)
equation:
\begin{equation}
  \ln {\rm E}[p(t)] = A + B(t_c-t)^m + C(t_c-t)^m\cos(\omega\ln (t_c-t) - \phi)
\label{eq:lppl}
\end{equation}
where $ B = - \kappa B' /m$ and $C = - \kappa C' / \sqrt{m^2+\omega^2}$. Note
that this expression (\ref{eq:lppl}) describes the average price dynamics only
up to the end of the bubble or negative bubble. The JLS model does not specify
what happens beyond $t_c$. This critical $t_c$ is the termination of the
bubble/negative bubble regime and the transition time to another regime. This
regime could be a big crash/rebound or a change of the growth rate of the
market.  The dynamics of the Merrill Lynch EMU (European Monetary Union)
Corporates Non-Financial Index in 2009 \cite{BFE-FCO09} provides a vivid
example of a change of regime characterized by a change of growth rate rather
than by a crash or rebound.  For $m<1$, the crash hazard rate accelerates up to
$t_c$ but its integral up to $t$, which controls the total probability for a
crash (or rebound) to occur up to $t$, remains finite and less than $1$ for all times $t
\leq t_c$. It is this property that makes it rational for investors to remain
invested knowing that a bubble (or negative bubble) is developing and that a crash (or rebound) is
looming. Indeed, there is still a finite probability that no crash or rebound
will occur during the lifetime of the bubble. The excess return $\mu(t) =
\kappa h(t)$ is the remuneration in the case of a bubble (the cost in the case
of a negative bubble) that investors require (accept to pay) to remain invested in
the bubbly asset, which is exposed to a crash (rebound0 risk. The condition that the
price remains finite at all time, including $t_c$, imposes that $m > 0$.

Within the JLS framework, a bubble is qualified when the hazard rate
accelerates. According to (\ref{eq:hazard}), this imposes $0<m<1$.  Since, by
definition, the hazard rate should be non-negative, an additional constraint
is~\cite{bm}
\begin{equation}
 b \equiv -Bm - |C|\sqrt{m^2+\omega^2}  \geq 0~,
  \label{eq:bg0}
\end{equation}
for bubbles and
\begin{equation}
 b \equiv -Bm - |C|\sqrt{m^2+\omega^2}  \leq 0~,
  \label{eq:le0}
\end{equation}
for negative bubbles.

\section{Prediction Method} \label{sec:pr}

We adapt the pattern recognition method of Gelfand et al. \cite{gel} to generate predictions
of crashes and rebound times in financial markets on the basis of the detection
and calibration of bubbles and negative bubbles. The prediction method used
here is basically that used to detect rebounds in \cite{rebound} but we now use
it to detect both crashes and rebounds at the same time.  Here we give a brief
summary of the method, which is decomposed into five steps.

\subsection{Fit the time series with the JLS model}\label{sec:fit}

Given a historical price time series of an index (such as the S\&P 500 or the
Dow Jones Industrial Average, for example), we first divide it into different
sub-windows $(t_1, t_2)$ of length $dt \equiv t_2 - t_1$ according to the
following rules:
\begin{enumerate}
\item The earliest start time of the windows is $t_{10}$.  Other
  start times $t_1$ are calculated using a step size of $dt_1 = 50$ calendar
  days.
\item The latest end time of the windows is $t_{20}$.  Other end
  times $t_2$ are calculated with a negative step size $dt_2 = -50$ calendar
  days.
\item The minimum window size $dt_{\mathrm{min}} = 110$ calendar days.
\item The maximum window size $dt_{\mathrm{max}} = 1500$ calendar days.
\end{enumerate}

For each sub-window generated by the above rules, the log of the index is fit
with the JLS equation (\ref{eq:lppl}). The fitting procedure is a combination
of a preliminary heuristic selection of the initial points and a local
minimizing algorithm (least squares). The linear parameters are slaved by the
nonlinear parameters before fitting. Details of the fitting algorithm can be
found in \cite{rebound,jls}.  We keep the best 10 parameter sets for each
sub-window and use these parameter sets as the input to the pattern recognition
method.

\subsection{Definition of crash and rebound}
\label{defreboundh2ysec}

We refer to a crash as `$\mathrm{Crh}$' and to a rebound as $ `\mathrm{Rbd}$'. A day
$d$ begins a crash (rebound) if the price on that day is the maximum (minimum)
price in a window of 100 days before and 100 days after. That is,
\begin{eqnarray}
  \mathrm{Crh} = \{d \mid P_d = \max\{P_x\}, \forall x \in
  [d-100,d+100]\}~\label{defcrashh2y}\\
  \mathrm{Rbd} = \{d \mid P_d = \min\{P_x\}, \forall x \in [d-100,d+100]\}~
  \label{defreboundh2y}
\end{eqnarray}
where $P_d$ is the adjusted closing price on day $d$. Our task is to diagnose
such crashes and rebounds in advance. We could also use other windows instead
of $\pm 100$ to define a rebound. The results are stable with respect to a
change of this number because we learn from the `learning set' with a certain
rebound window width and then try to predict the rebounds using the same window
definition. The reference \cite{rebound} shows the results for $\pm 200$-days
and $\pm 365$-days type of rebounds.

\subsection{Learning set, class, group and informative parameter}

As described above, we obtain a set of parameters that best fit the model
(\ref{eq:lppl}) for each window. Then we select a subset of the whole set which
only contains the fits of crashes and rebounds with critical times found within the window
(that is, where parameter $t_c$ is \emph{not} calculate to be beyond the window
bounds). We learn the properties of historical rebounds from this set and
develop the predictions based on these properties. We call this set the
\emph{learning set}. In this paper, a specific day for each index is chosen as
the `present time' (for backtesting purposes). All the fit windows before that
day will be used as the learning set and all the fit windows after that will be
used as the \emph{testing set}, in which we will predict future rebounds. The
quality of the predictability of this method can be quantified by studying the
predicted results in the testing set using only the information found in the
learning set.

Each of the sub-windows generated by the rules in Sec. \ref{sec:fit} will be
assigned one of two \emph{classes} and one of 14 \emph{groups}. Classes
indicate how close the modeled critical time $t_c$ is to a historical crash or
rebound, where Class I indicates `close' and Class II indicates `not close'
(`close' will be defined below as a parameter).  Groups of windows have similar
window widths. For each fit, we create a set of six parameters: $m, \omega,
\phi$ and $B$ from Eq.~(\ref{eq:lppl}), $b$ from Eq.~(\ref{eq:bg0}) and $q$ as
the residual of the fit.  We will compare the probability density functions (pdf) of these parameters among
the different classes and groups.  The main goal of this technique is \emph{to
identify patterns of parameter pdf's that are different between windows with
crashes or rebounds and windows without crashes or rebounds.}  Given such a difference and a new,
out-of-sample window, we can probabilistically state that a given window will
or will not end in a crash or rebound.

\begin{figure}[htb]
\centering
\includegraphics[width=0.9\textwidth]{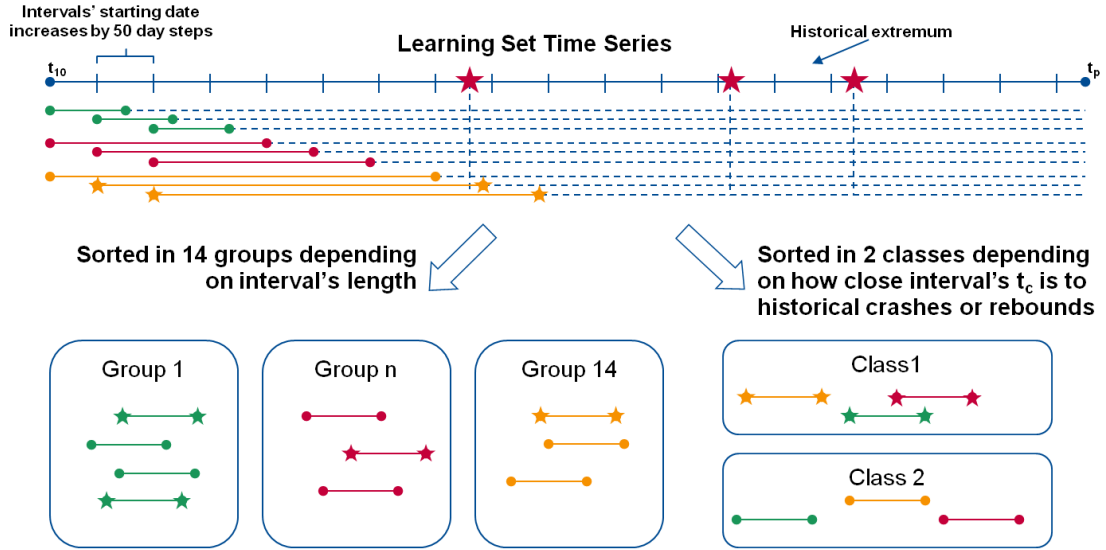}
\caption{{\bf Step one of pattern recognition procedure:} Create sub-windows,
fit each window with the JLS model. Classify the fits in groups and classes.}
\label{fig:pr1}
\end{figure}

\begin{figure}[htb]
\centering
\includegraphics[width=0.9\textwidth]{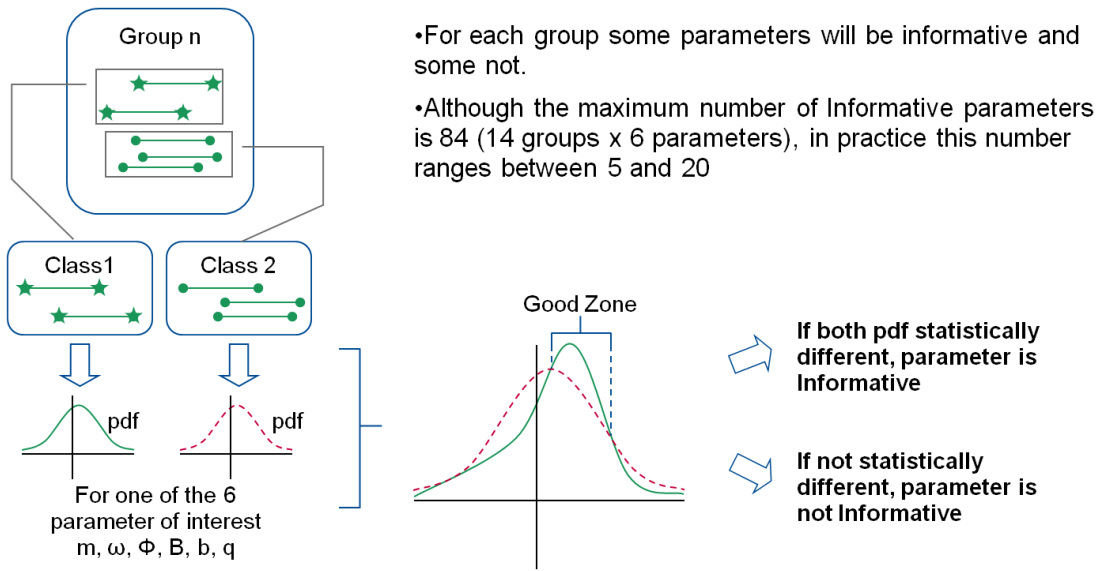}
\caption{{\bf Step two of pattern recognition procedure:} For each group
compare fits in class I with those in class II and find out the informative
parameters.} \label{fig:pr2}
\end{figure}

A figure is very helpful for understanding these concepts. We show the
selection of the sub-windows and sort the fits by classes and groups in
Fig.~\ref{fig:pr1}. Then we create the pdf's of
each of these parameters for each fit and define \emph{informative parameters}
as those parameters for which the pdf's differ significantly according to a
Kolmogorov-Smirnov test. For each informative parameter, we find the regions of
the abscissa of the pdf for which the Class I pdf (fits with $t_c$ close to an
extremum) is greater than the Class II pdf. This procedure has been performed
in Fig.~\ref{fig:pr2}.

[Figs.~\ref{fig:pr1}--\ref{fig:pr2} about here.]

\subsection{Questionnaires, traits and features}
\label{sec:prqtf}

Using the informative parameters and their pdf's described above, we can
generate a \emph{questionnaire} for each day of the learning or testing set.  A
questionnaire is a quantitative inquiry into whether or not a set of parameters
is likely to indicate a bubble or negative bubble. Questionnaires will be used
to identify bubbles (negative bubbles) which will be followed by crashes
(rebounds).  In short, the length of a questionnaire tells how many informative
parameters there are for a given window size.  An informative parameter implies
that the pdf's of that parameter are very different for windows with a
crash/rebound than windows without.  The more values of `1' in the
questionnaire, the more likely it is that the parameters are associated with a
crash/rebound.

\begin{figure}[htb]
\centering
\includegraphics[width=0.9\textwidth]{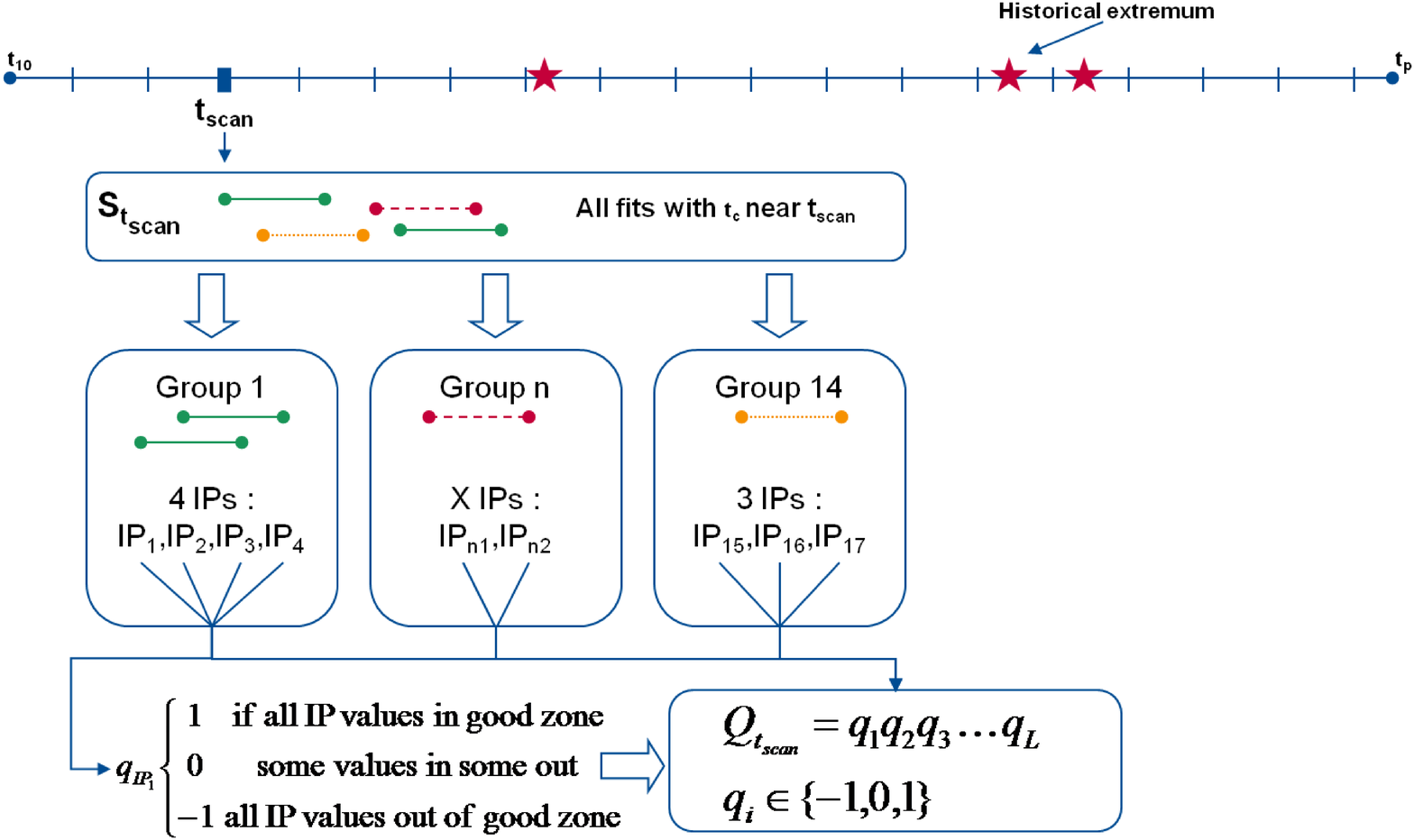}
\caption{{\bf Step three of pattern recognition procedure:} Generate the
questionnaire for each trading day.} \label{fig:pr3}
\end{figure}

One questionnaire is constructed for each day $t_{scan}$ in our learning set.
We first collect all the fits which have a critical time near that day (`near'
will be defined). Then we create a string of bits whose length is equal to the
number of informative parameters found.  Each bit can take a value -1, 0 or 1
(a balanced ternary system).  Each bit represents the answer to the question:
are more than half of the collected fits more likely to be considered as Class
I? If the answer is `yes', we assign $1$ in the bit of the questionnaire
corresponding to this informative parameter.  Otherwise, we assign $0$ when the
answer cannot be determined or $-1$ when the answer is `no'.  A visual
representation of this questionnaire process is shown in Fig.~\ref{fig:pr3}.

[Fig.~\ref{fig:pr3} about here.]

The concept of a \emph{trait} is developed to describe the property of the
questionnaire for each $t_{scan}$. Each questionnaire can be decomposed into a
fixed number of traits if the length of the questionnaire is fixed. We will not
give the details in how the traits are generated in this paper. For a clear
explanation of the method, please refer to Sec. 3.9 of the reference
\cite{rebound}.  Think of a trait as a sub-set (like the `important' short
section from a very long DNA sequence) of a fixed-length questionnaire that is
usually found in windows that show crashes/rebounds.  Conversely, a trait can
indicate windows where a crash/rebound is not found.

Assume that there are two sets of traits $T_I$ and $T_{II}$ corresponding to
Class I and Class II, respectively.  Scan day by day the date $t$ before the
last day of the learning set. If $t$ is `near' an extreme event (crash or
rebound), then all traits generated by the questionnaire for this date belong
to $T_I$. Otherwise, all traits generated by this questionnaire belong to
$T_{II}$. `Near' is defined as at most 20 days away from an extreme event. The
same definition will be used later in Section~\ref{sec:prediction} when we
introduce the error diagram.

Using this threshold, we declare that an alarm starts on the first day that the
unsorted crash alarm index time series exceeds this threshold. The duration of
this alarm $D_a$ is set to 41 days, since the longest distance between a crash
and the day with index greater than the threshold is 20 days.

\begin{figure}[htb] \centering
\includegraphics[width=0.9\textwidth]{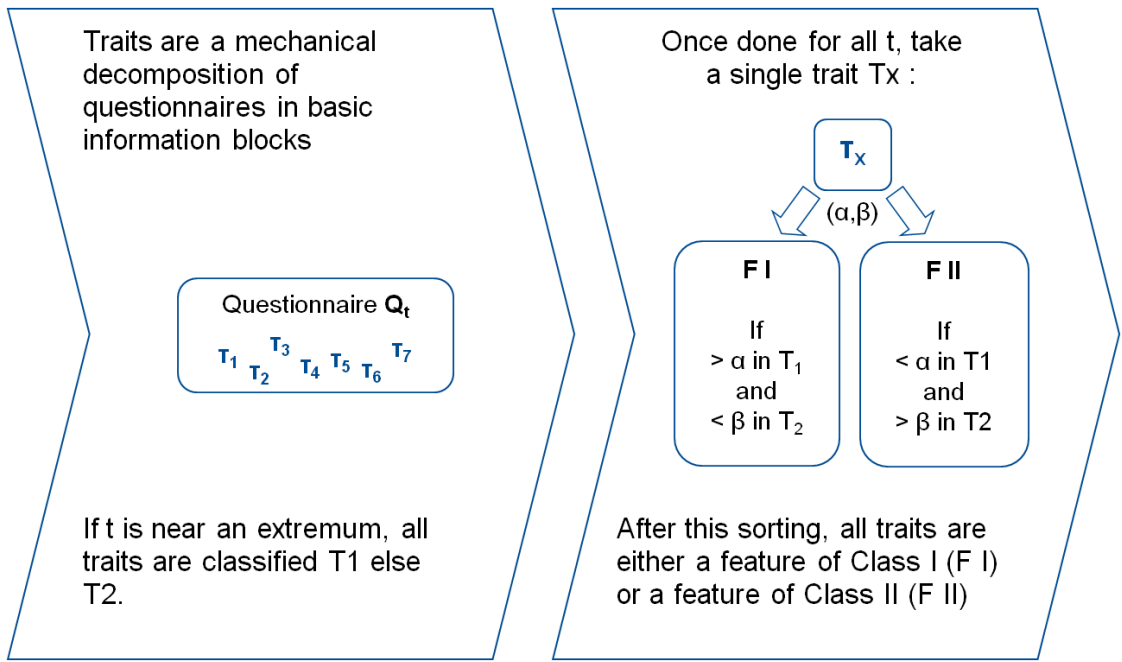}
\caption{{\bf Step four of pattern recognition procedure:} Create traits from
questionnaires. Obtain features for each class by traits statistics.}
\label{fig:pr4}
\end{figure}

Count the frequencies of a single trait $\tau$ in $T_I$ and $T_{II}$. If $\tau$
is in $T_I$ for more than $\alpha$ times and in $T_{II}$ for less than $\beta$
times, then we call this trait $\tau$ a \emph{feature} $F_I$ of Class I.
Similarly, if $\tau$ is in $T_{I}$ for less than $\alpha$ times and in $T_{II}$
for more than $\beta$ times, then we call $\tau$ a \emph{feature} $F_{II}$ of
Class II.  The pair $(\alpha, \beta)$ is defined as a \emph{feature
qualification}. Fig.~\ref{fig:pr4} shows the generation process of traits and
features. We would like to clarify that by definition some of the traits are
not from any type of feature since they are not `extreme' and we cannot extract
clear information from them.

[Fig.~\ref{fig:pr4} about here.]

\subsection{Alarm index}

\begin{figure}[htb]
\centering
\includegraphics[width=0.9\textwidth]{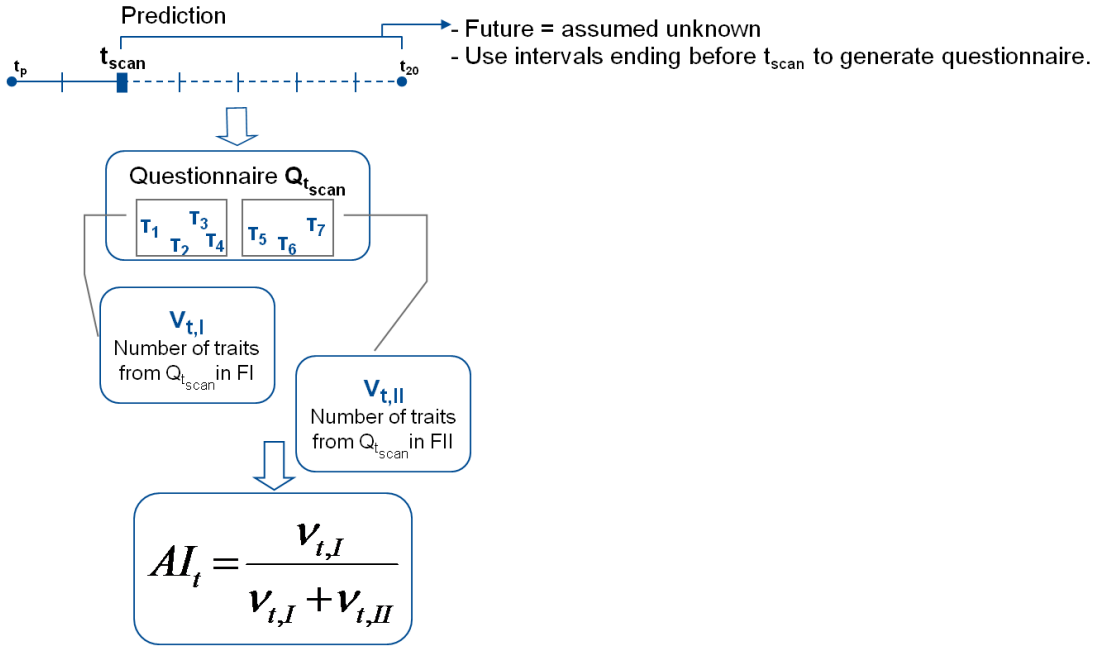}
\caption{{\bf Step five of pattern recognition procedure:} Construct the alarm
index for each day by decomposing the questionnaire into traits for that day
and compare these traits to the features of each class.} \label{fig:pr5}
\end{figure}

The final piece in our methodology is to define an \emph{alarm index} for both
crashes and rebounds. An alarm index is developed based on features to show the
probability that a certain day is considered to be a rebound or a crash. We
first collect all the fits which have a predicted critical time near this
specific day and generate questionnaires and traits from these fits. The
rebound (crash) alarm index for a certain day is just a ratio quantified by the
total number of traits from feature type $F_I$ (a set of traits which have high
probability to represent rebound (crash)) divided by the total number of traits
from both $F_I$ and $F_{II}$. Note that $F_{II}$ is a set of traits which have
low probability to represent rebound (crash). The principles for the generation of the alarm index
are summarized in Fig.~\ref{fig:pr5}.

[Fig.~\ref{fig:pr5} about here.]

Two types of alarm index are developed. One is for the back tests in the
learning set, as we have already used the information before this time to
generate informative parameters and features. The other alarm index is for the
prediction tests.  We generate this prediction alarm index using only the
information before a certain time and then try to predict crashes and rebounds
in the `future'.

\section{Prediction in major equity markets} \label{sec:prediction}
We perform systematic detections and forecasts on both the market crashes and
rebounds for 10 major global equity markets using the method we discussed in
Sec.~\ref{sec:pr}. The 10 indices are S\&P 500 (US), Nasdaq composite (US),
Russell 2000 (US), FTSE 100 (UK), CAC 40 (France), SMI (Switzerland), DAX
(German), Nikkei 225 (Japan), Hang Seng (Hong Kong) and ASX (Australia). The
basic information for these indices used in this study is listed in
Tab.~\ref{tab:numintervals}. Due to space constraints, we cannot show all
results for these 10 indices here. The complete results for three indices from
different continents are shown in this paper. They are Russell 2000 (America),
SMI (Europe) and Nikkei 225 (Asia). Partial results for the remaining indices
are shown in Tab.~\ref{tab:allstratsharpes}, which will be discussed in details
later in Sec.~\ref{sec:strategy}.

The alarm index depends on the features which are generated using information
from the learning set.  Thus, the alarm index before the end of the learning
set uses `future' information.  That is, the value of the alarm index on a
certain day $t_0$ in the learning set uses prices found at $t > t_0$ to
generate features. The feature definitions from the learning set are then used
to define the alarm index in the \emph{testing} set using \emph{only} past
prices.  That is, the value of the alarm index on a certain day $t_0$ in the
\emph{testing} set uses only prices found at $t \leq t_0$ in the testing set
(and the \emph{definitions} of features found in the \emph{learning} set).  We
do not use `future' information in the testing set.  In this case, the alarm
index predicts crashes and rebounds in the market.

\begin{figure}[htb]
\centering
\includegraphics[width=0.9\textwidth]{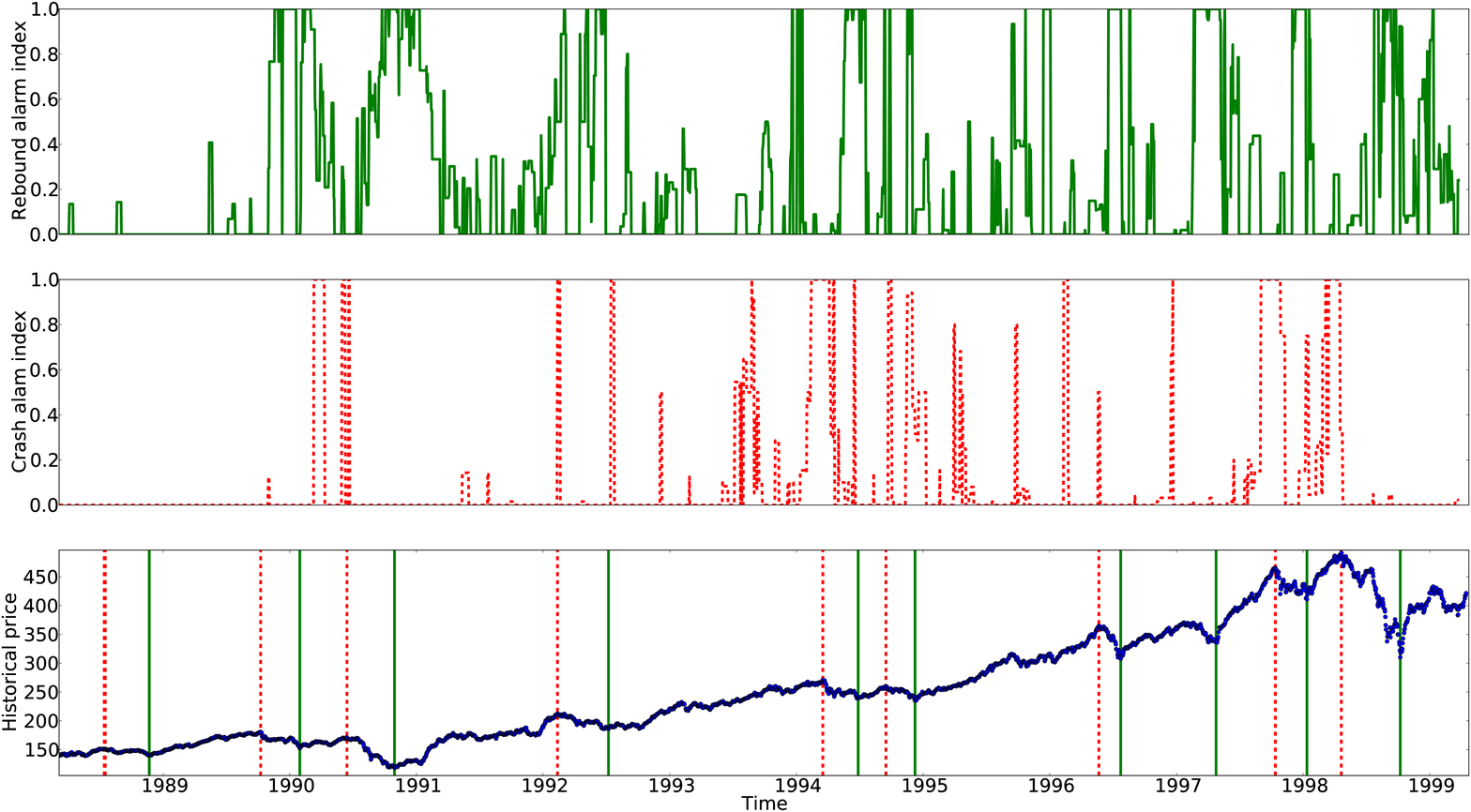}
\caption{{\bf Alarm index and log-price of the Russell 2000 Index for the
\emph{learning} set}, where the end date is 17-Apr-1999. (upper) Rebound alarm
index for the learning set using feature qualification pair $(7, 100)$. The
rebound alarm index is in the range $[0,1]$. The higher the rebound alarm
index, the more likely is the occurrence of a rebound. (middle) Crash alarm
index for the learning set using feature qualification pair $(7, 100)$. The
crash alarm index is in the range $[0,1]$. The higher the crash alarm index,
the more likely is the occurrence of a crash. (lower) Plot of price versus time
of Russell Index (shown in blue cycles). Green solid vertical lines indicate
rebounds defined by local minima within plus and minus 100 days around them.
Red dashed vertical lines indicate crashes defined by local maxima within plus
and minus 100 days around them. Note that these rebounds and crashes are the
historical ``change of regime'' rather than only the jump-like reversals.}
\label{fig:russellb}
\end{figure}

\begin{figure}[htb]
\centering
\includegraphics[width=0.9\textwidth]{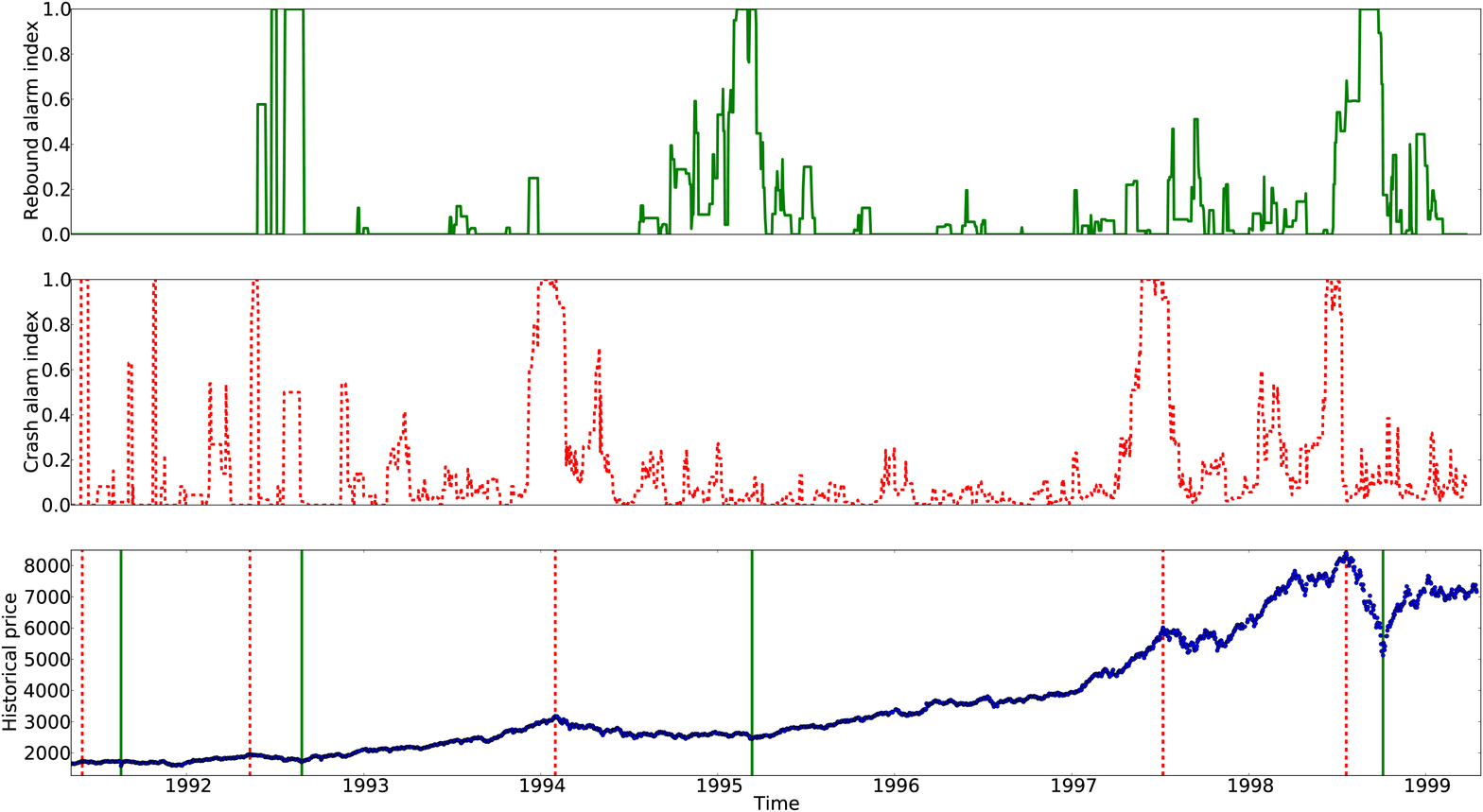}
\caption{{\bf Alarm index and price of the SMI Index for the \emph{learning}
set}, where the end date is 17-Apr-1999. The format is the same as
Fig.~\protect\ref{fig:russellb}.}  \label{fig:smib}
\end{figure}

\begin{figure}[htb]
\centering
\includegraphics[width=0.9\textwidth]{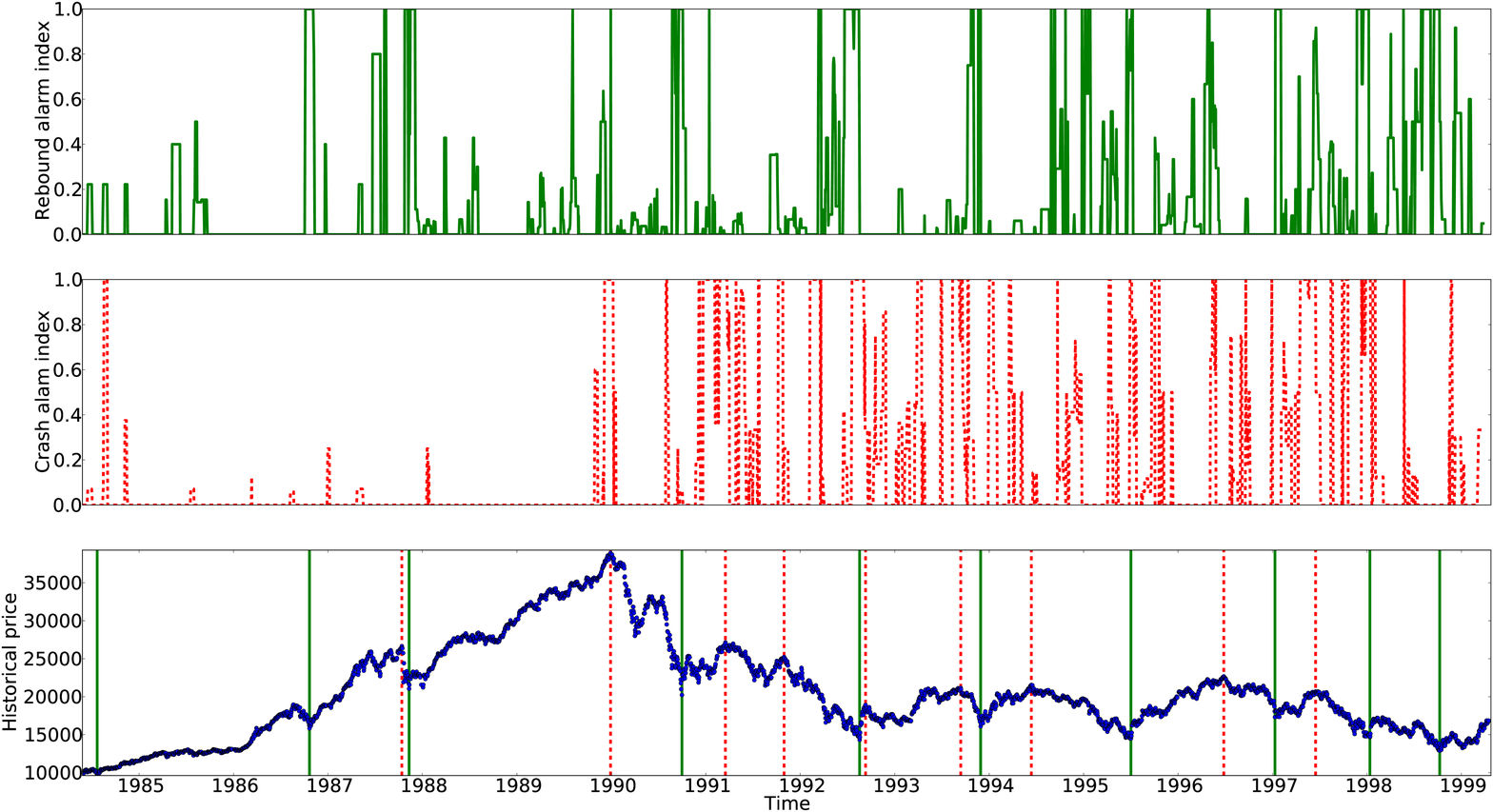}
\caption{{\bf Alarm index and price of the Nikkei 225 Index for the
\emph{learning} set}, where the end date is 17-Apr-1999. The format is the same
as Fig.~\ref{fig:russellb}. } \label{fig:nikkeib}
\end{figure}

\begin{figure}[htb]
\centering
\includegraphics[width=0.9\textwidth]{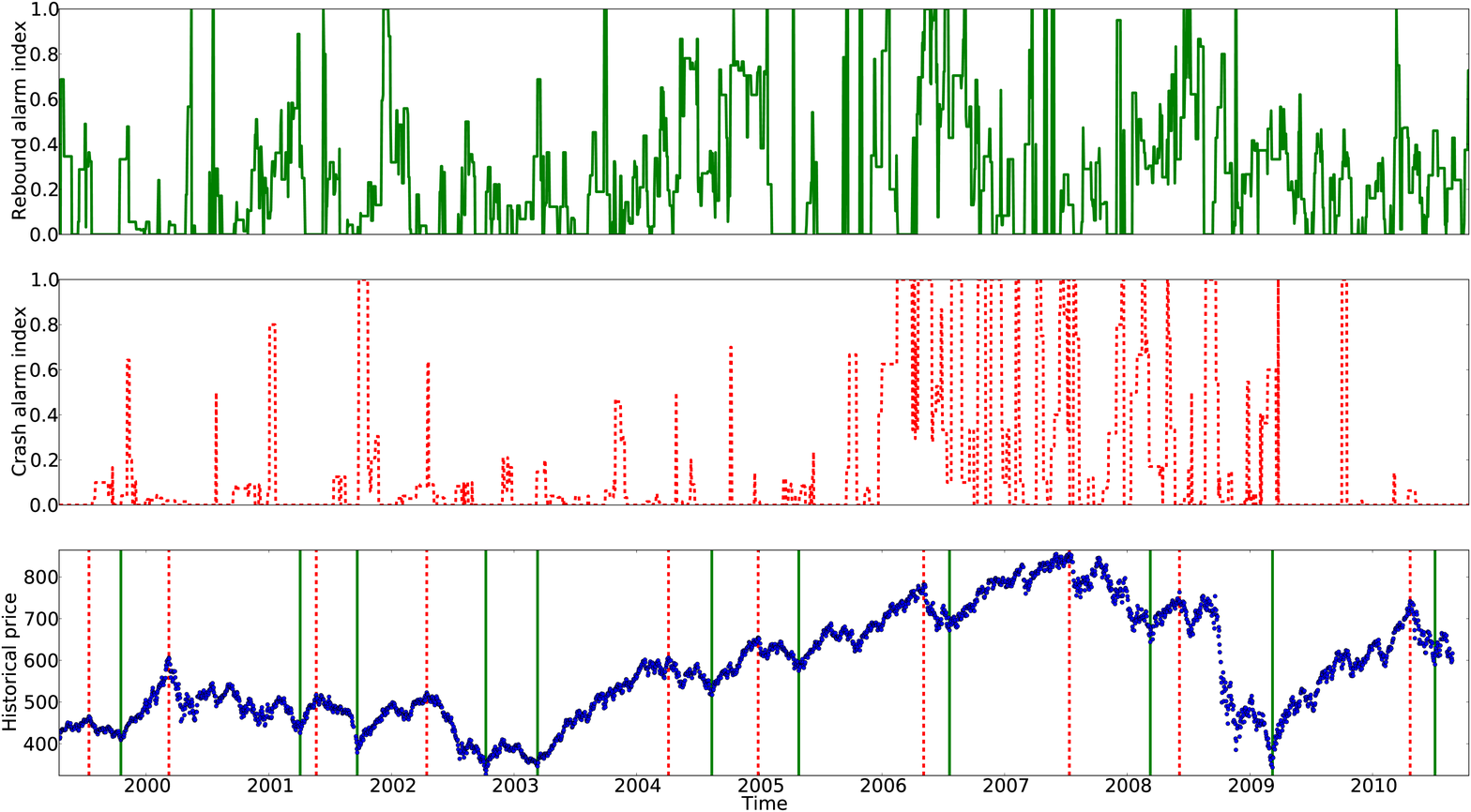}
\caption{{\bf Alarm index and log-price of Russell 2000 Index for the
\emph{predicting} set} after 17-Apr-1999. The format is the same as
Fig.~\ref{fig:russellb}.} \label{fig:russellp}
\end{figure}

\begin{figure}[htb]
\centering
\includegraphics[width=0.9\textwidth]{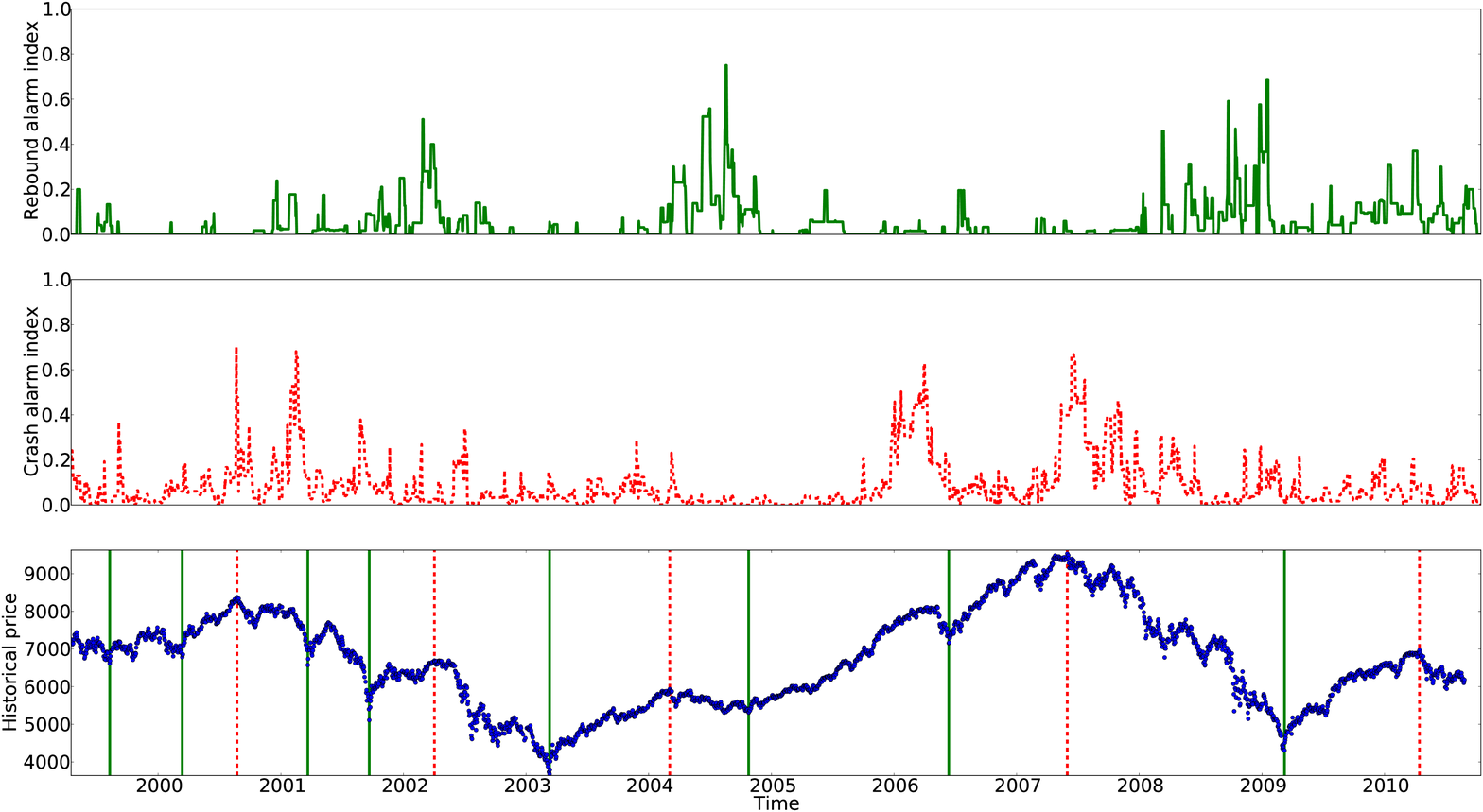}
\caption{{\bf Alarm index and price of the SMI Index for the \emph{predicting}
set} after 17-Apr-1999. The format is the same as Fig.~\ref{fig:russellb}. }
\label{fig:smip}
\end{figure}

\begin{figure}[htb]
\centering
\includegraphics[width=0.9\textwidth]{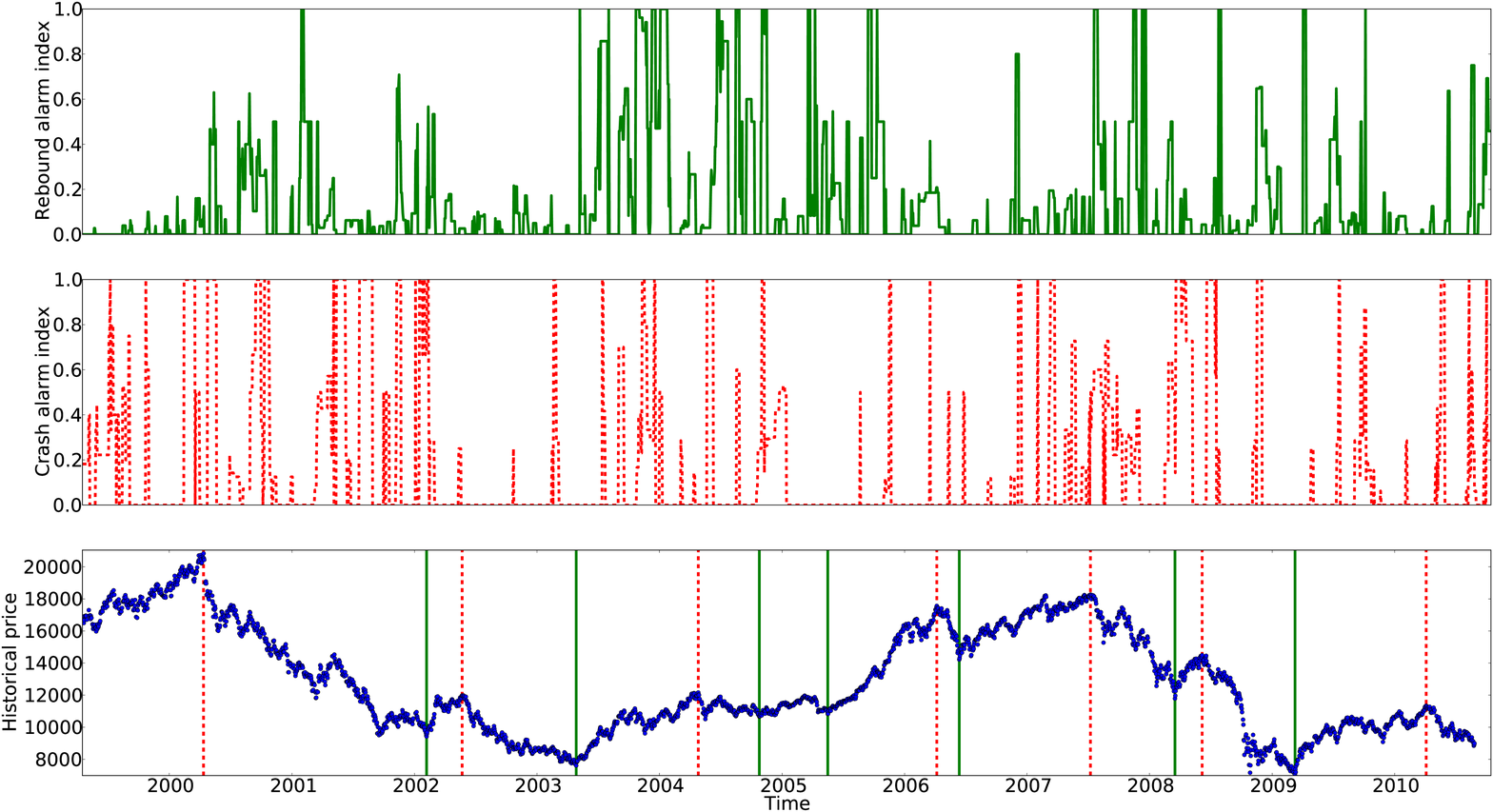}
\caption{{\bf Alarm index and price of the Nikkei 225 Index for the
\emph{predicting} set} after 17-Apr-1999. The format is the same as
Fig.~\ref{fig:russellb}. } \label{fig:nikkeip}
\end{figure}

The crash and the rebound alarm index for Russell 2000, SMI and Nikkei are
shown in Figs.~\ref{fig:russellb}--\ref{fig:nikkeip}.
Figs.~\ref{fig:russellb}--\ref{fig:nikkeib} show the back testing results for
these three indices and Figs.~\ref{fig:russellp}--\ref{fig:nikkeip} present the
prediction results. In all of these results, the feature qualification pair
$(7, 100)$ is used. This means that a certain trait must appear in trait Class
I (crash or rebound) at least 7 times \emph{and} must appear in trait Class II
(no crash or rebound) less than 100 times.  If so, then we say that this trait
is a feature of Class I. If, on the other hand, the trait appears 7 times or
less in Class I \emph{or} appears 100 times or more in Class II, then this
trait is a feature of Class II.  Tests on other feature qualification pairs are
performed also. Due to the space constraints, we do not show the alarm index
constructed by other feature qualification pairs here, but later we will
present the predictability of these alarm indices by showing the corresponding
error diagrams. In the rest of this paper, if we do not mention otherwise, we
use $(7, 100)$ as the feature qualification pair.

[Figs.~\ref{fig:russellb}--\ref{fig:nikkeip} about here.]

To check the quality of the alarm index quantitatively, we introduce error
diagrams \cite{Molchan1,Molchan2}.  Using Nikkei 225 as an example, we create
an error diagram for crash predictions after 17-Apr-1999 with a certain feature
qualification in the following way:
\begin{enumerate}
\item Calculate features and define the alarm index using the \emph{learning
    set} between 04-Jan-1984 and 17-Apr-1999.
\item Count the number of crashes after 17-Apr-1999 as defined in
  Sec. \ref{defreboundh2ysec} and expression (\ref{defcrashh2y}). There are 7
  crashes.
\item Take the crash alarm index time series (after 17-Apr-1999) and sort the set
of all alarm index values in decreasing order. There are 4,141
points in this series and the sorting operation delivers a list of
4,141 index values, from the largest to the smallest one.
\item  The largest value of this sorted series defines the first
  threshold.
\item Using this threshold, we declare that an alarm starts on the first day
  that the unsorted crash alarm index time series exceeds this threshold.  The
  duration of this alarm $D_a$ is set to 41 days, since the longest distance
  between a crash and the day with index greater than the threshold is 20 days.
  This threshold is consistent with the previous classification of
  questionnaires in Section~\ref{sec:prqtf}, where we define a predicted
  critical time as `near' the real extreme events when its distance is less
  than 20 days. Then, a prediction is deemed successful when a crash
  falls inside that window of 41 days.
\item If there are no successful predictions at this threshold, move the
  threshold down to the next value in the sorted series of alarm index.
\item Once a crash is predicted with a new value of the threshold, count the
  ratio of unpredicted crashes (unpredicted crashes / total crashes in set)
  and the ratio of alarms used (duration of alarm period / 4,141 prediction
  days). Mark this as a single point in the error diagram.
\end{enumerate}
In this way, we will mark 7 points in the error diagram for the 7 Nikkei 225
crashes after 17-Apr-1999.

The aim of using such an error diagram in general is to show that a given
prediction scheme performs better than random.  A random prediction follows the
line $y = 1 - x$ in the error diagram. A set of points below this line
indicates that the prediction is better than randomly choosing alarms. The
prediction is seen to improve as more error diagram points are found near the
origin point $(0, 0)$. The advantage of error diagrams is to avoid discussing
how different observers would rate the quality of predictions in terms of the
relative importance of avoiding the occurrence of false positive alarms and of
false negative missed rebounds. By presenting the full error diagram, we thus
sample all possible preferences and the unique criterion is that the error
diagram curve be shown to be statistically significantly below the
anti-diagonal $y = 1 - x$.

\begin{figure}[htbp]
\centering
\includegraphics[width=0.49\textwidth]{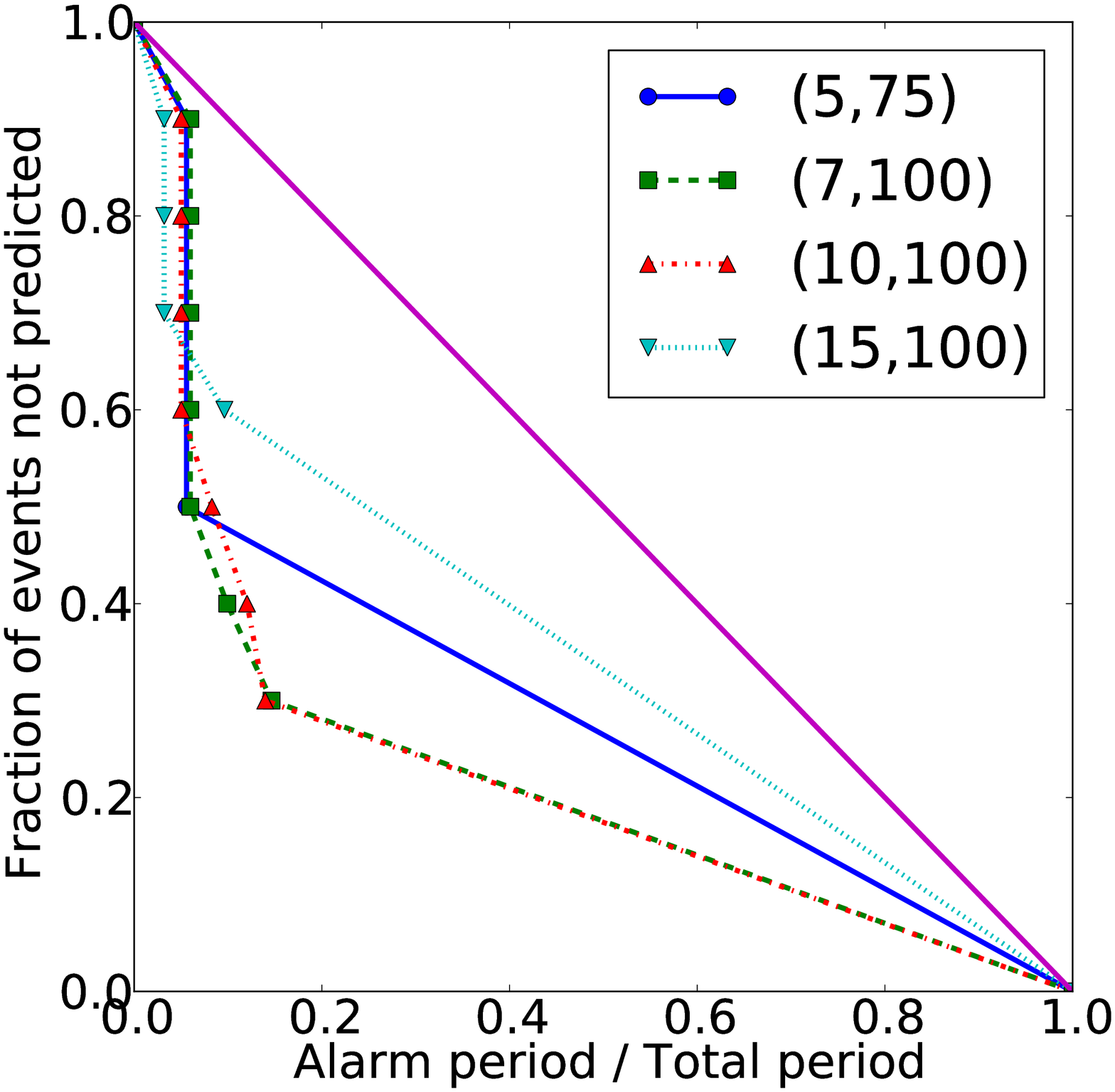}
\includegraphics[width=0.49\textwidth]{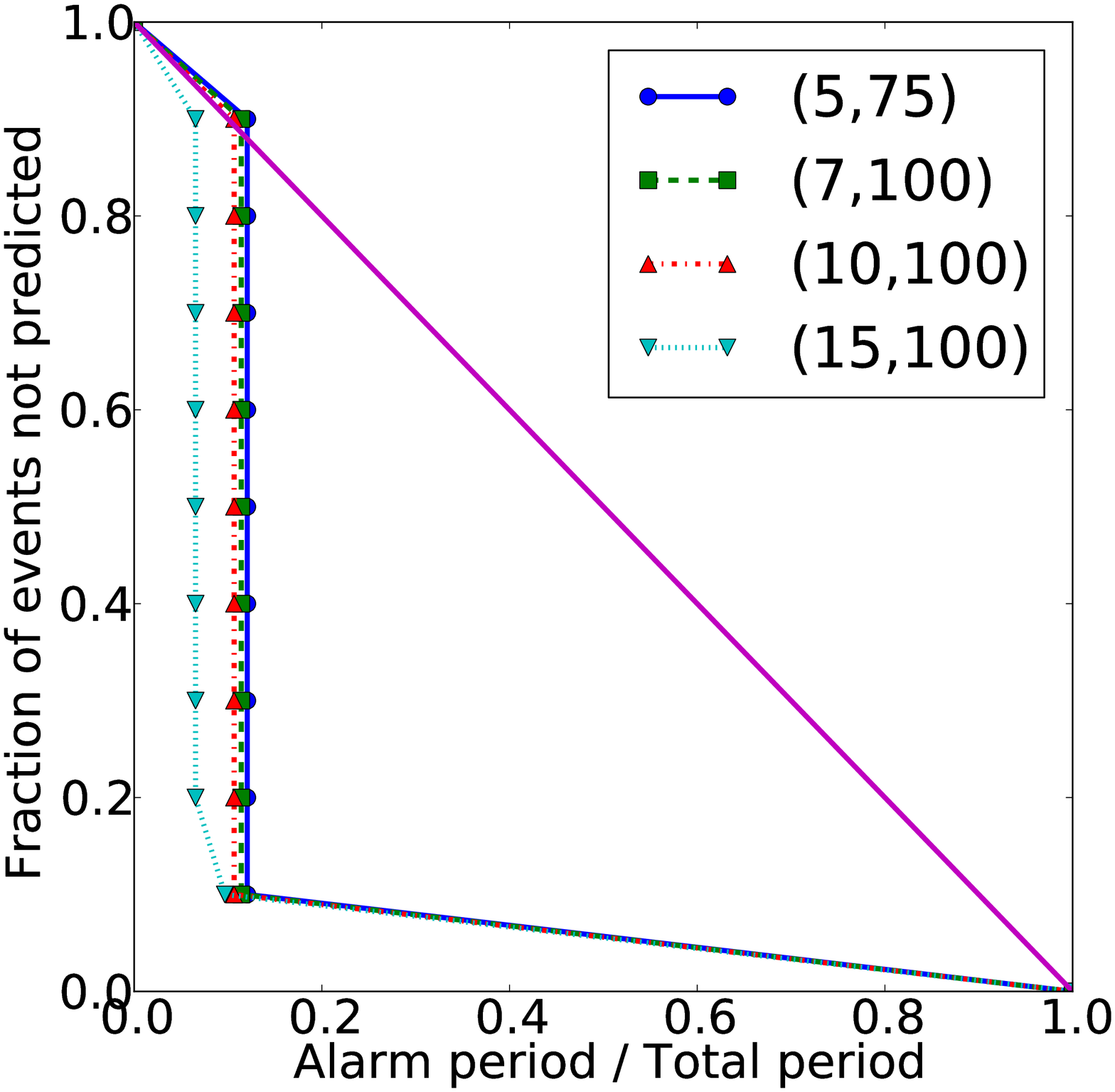}
\includegraphics[width=0.49\textwidth]{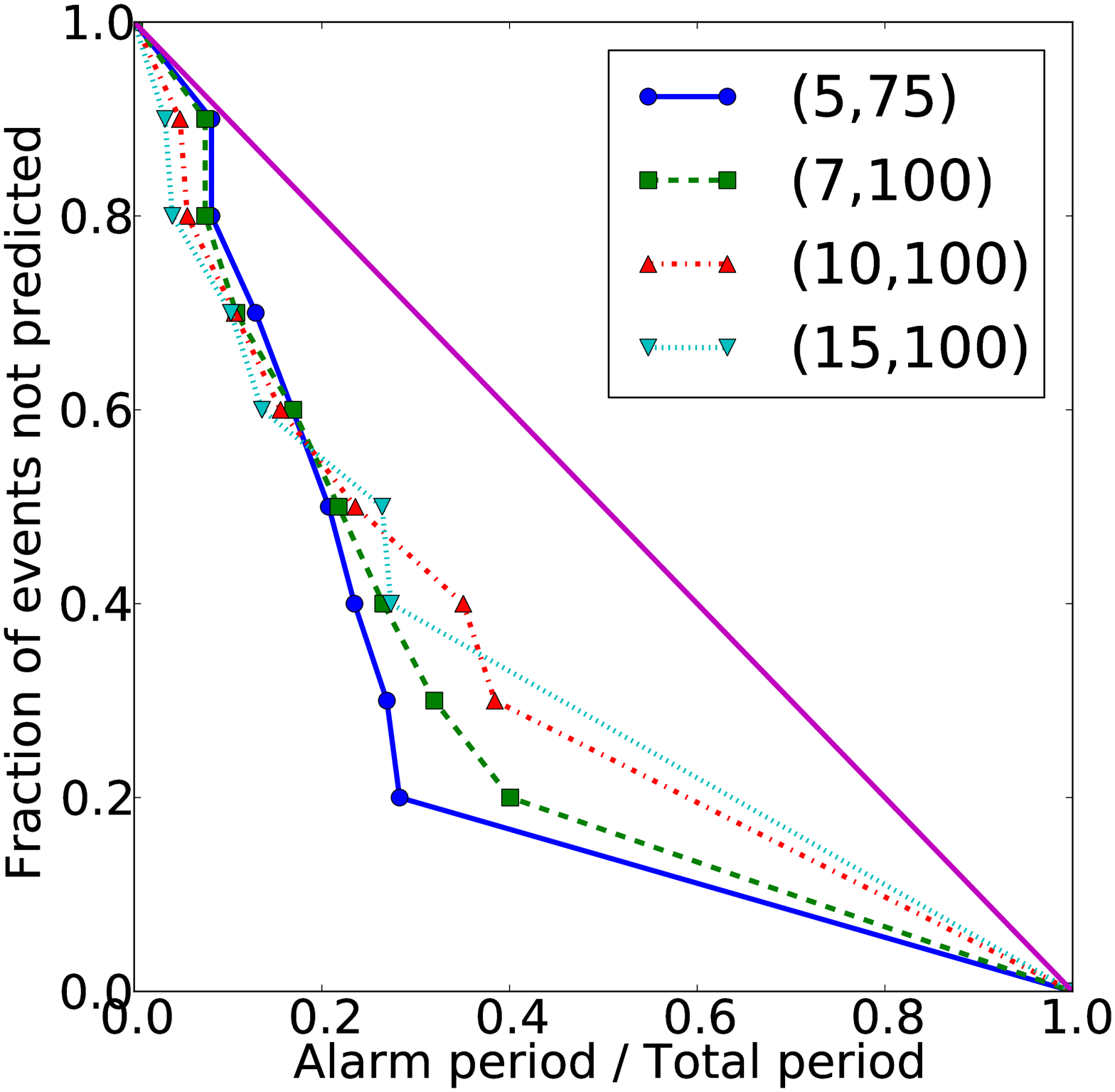}
\includegraphics[width=0.49\textwidth]{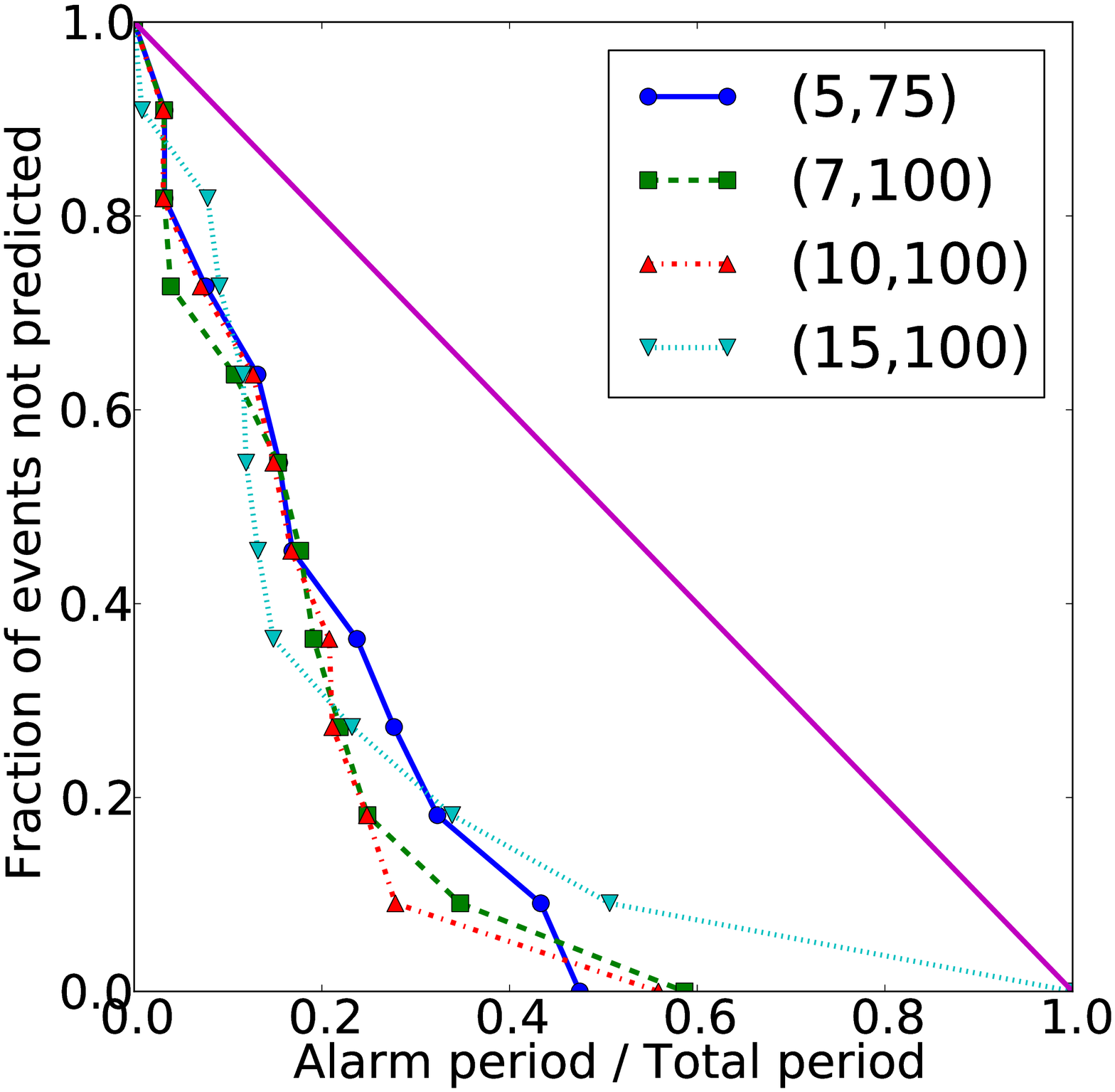}
\caption{{\bf Error diagram for back tests and predictions of crashes and
rebounds for Russell 2000 index with different types of feature
qualifications.} The value of the feature qualifications are shown in the
legend. The fact that all the curves lie under the line $y=1-x$ indicates
better performance than chance in detecting crashes and rebounds using our method.\\
(Upper left) Back tests of crashes. (Upper right) Back tests of rebounds.
(Lower left) Predictions of crashes. (Lower right) Predictions of rebounds.
Feature qualification $(\alpha, \beta)$ means that, if the occurrence of a
certain trait in Class I is larger than $\alpha$ and less than $\beta$, then we
call this trait a feature of Class I and vice versa. See text for more
information.} \label{fig:errrussell}
\end{figure}

\begin{figure}[htbp]
\centering
\includegraphics[width=0.49\textwidth]{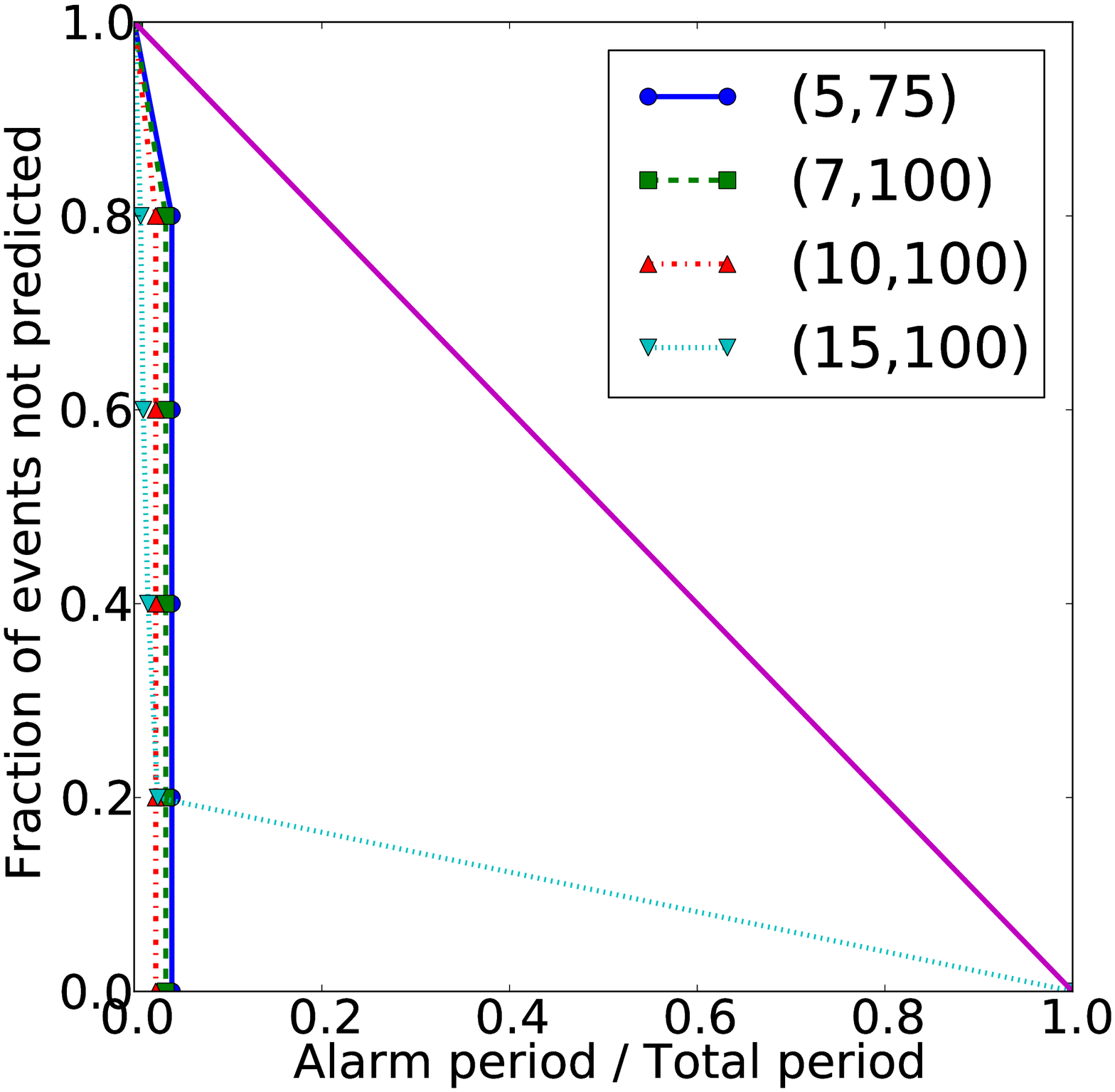}
\includegraphics[width=0.49\textwidth]{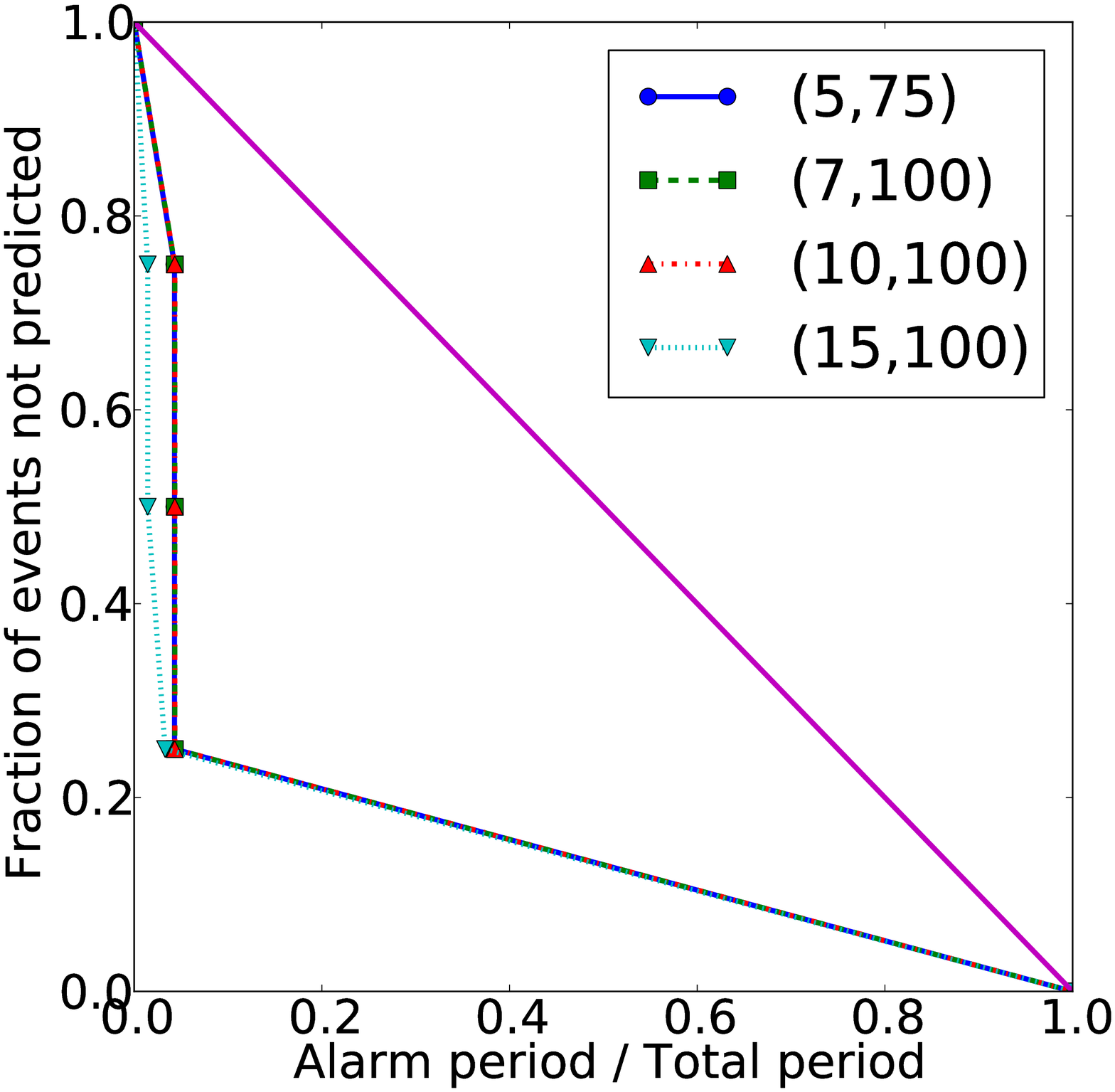}
\includegraphics[width=0.49\textwidth]{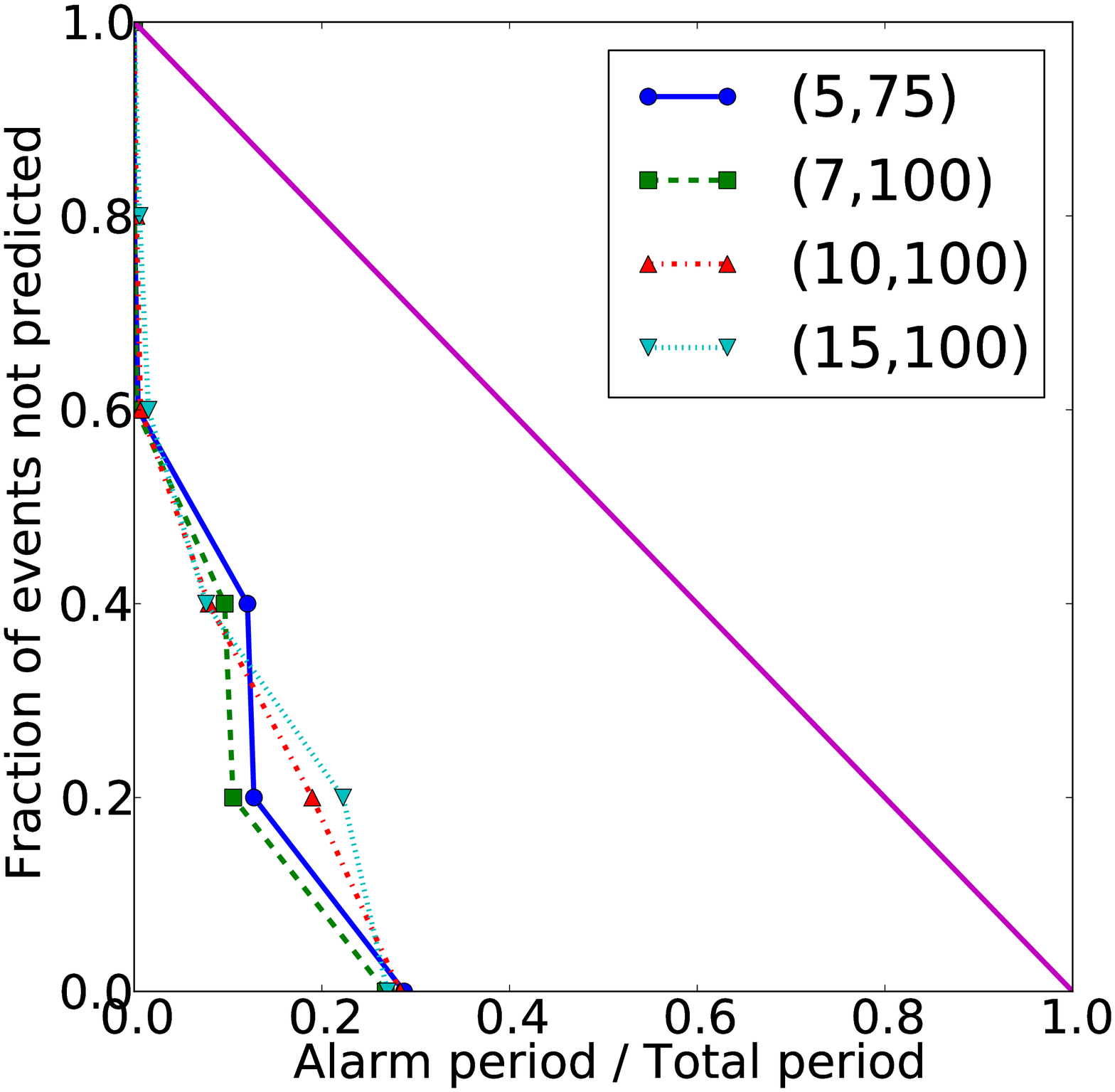}
\includegraphics[width=0.49\textwidth]{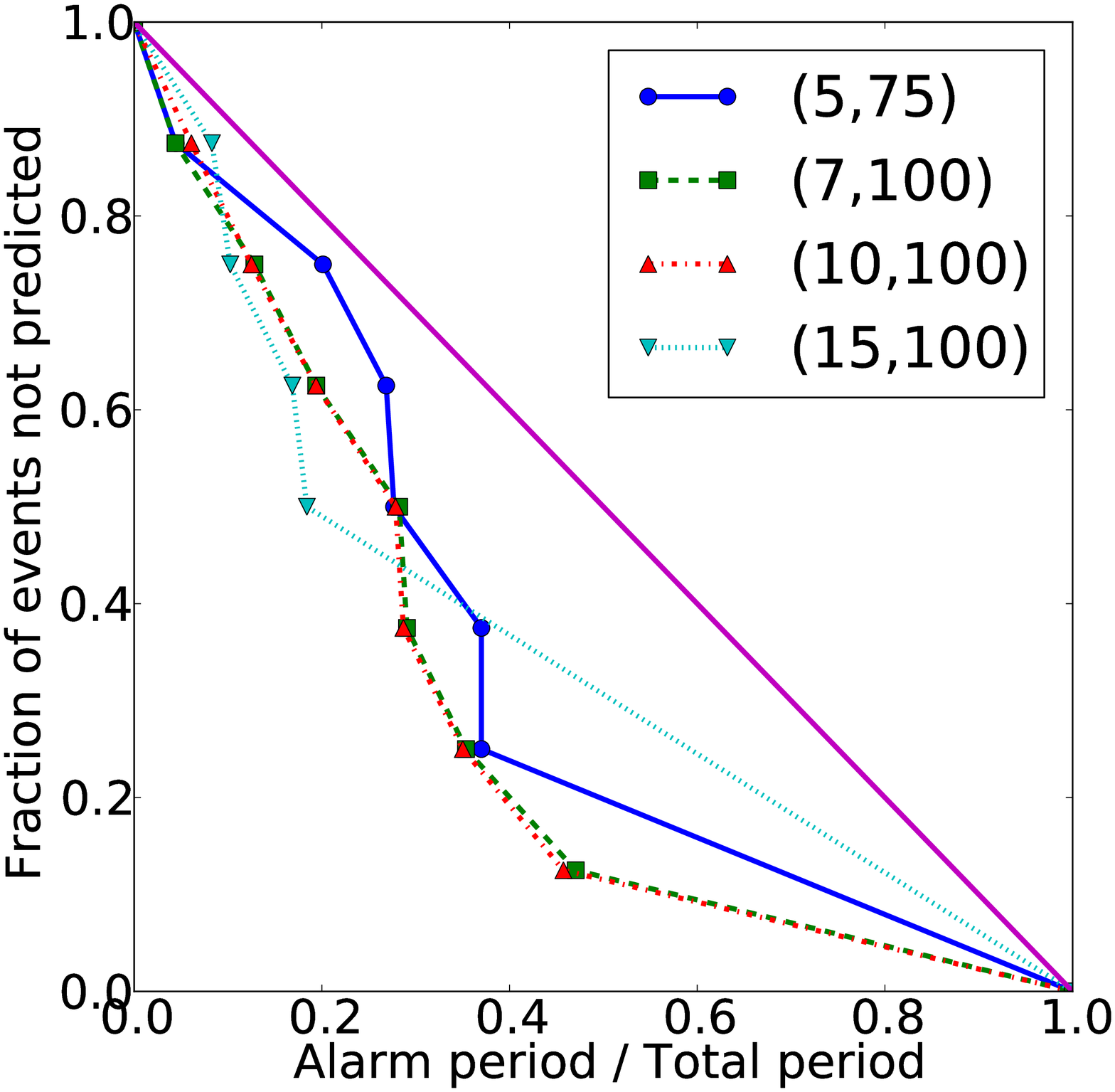}
\caption{{\bf Error diagram for back tests and predictions of crashes and
rebounds for SMI index with different types of feature qualifications.} The
format is the same as Fig.~\ref{fig:errrussell}.} \label{fig:errsmi}
\end{figure}

\begin{figure}[htbp]
\centering
\includegraphics[width=0.49\textwidth]{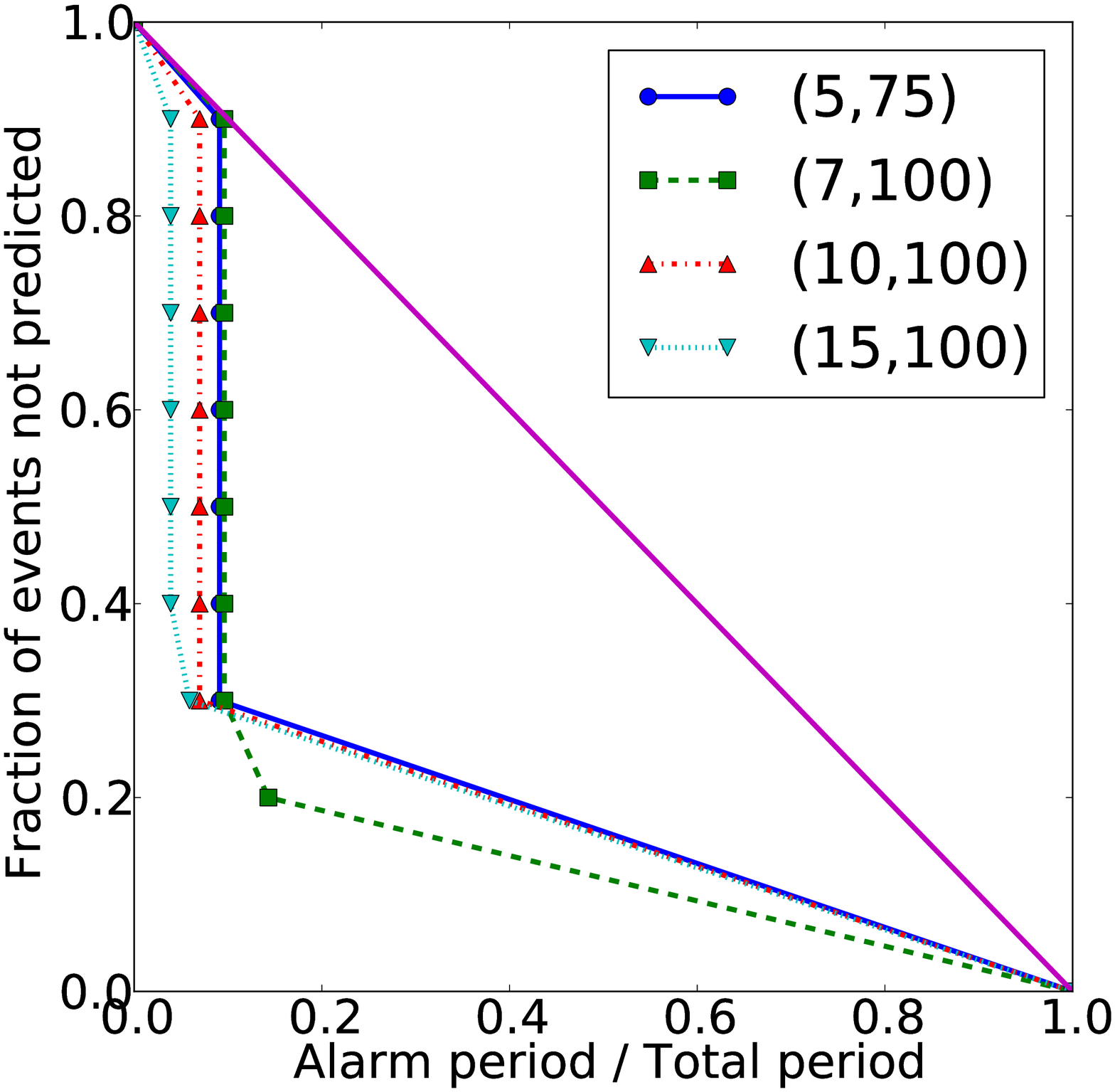}
\includegraphics[width=0.49\textwidth]{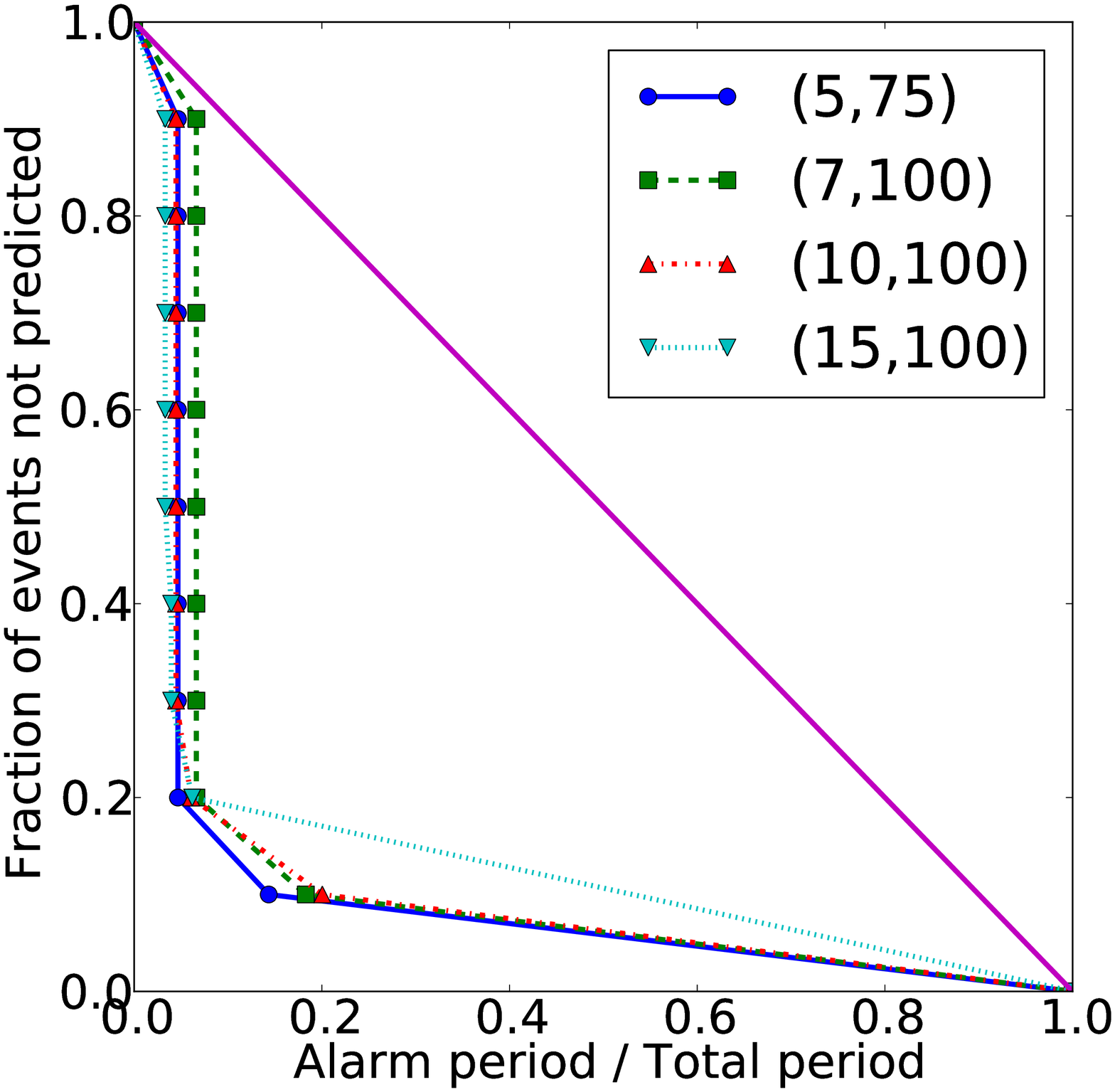}
\includegraphics[width=0.49\textwidth]{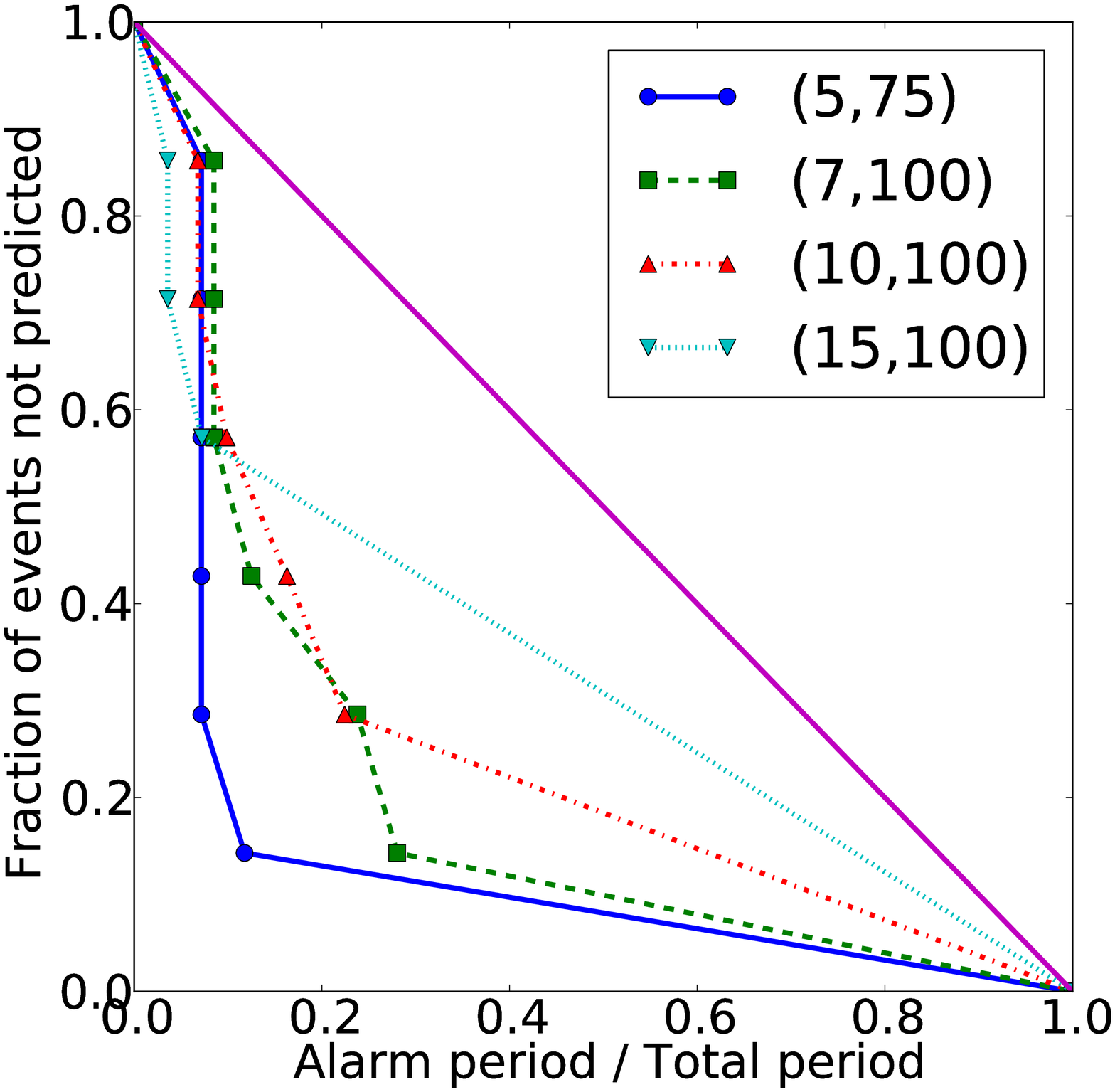}
\includegraphics[width=0.49\textwidth]{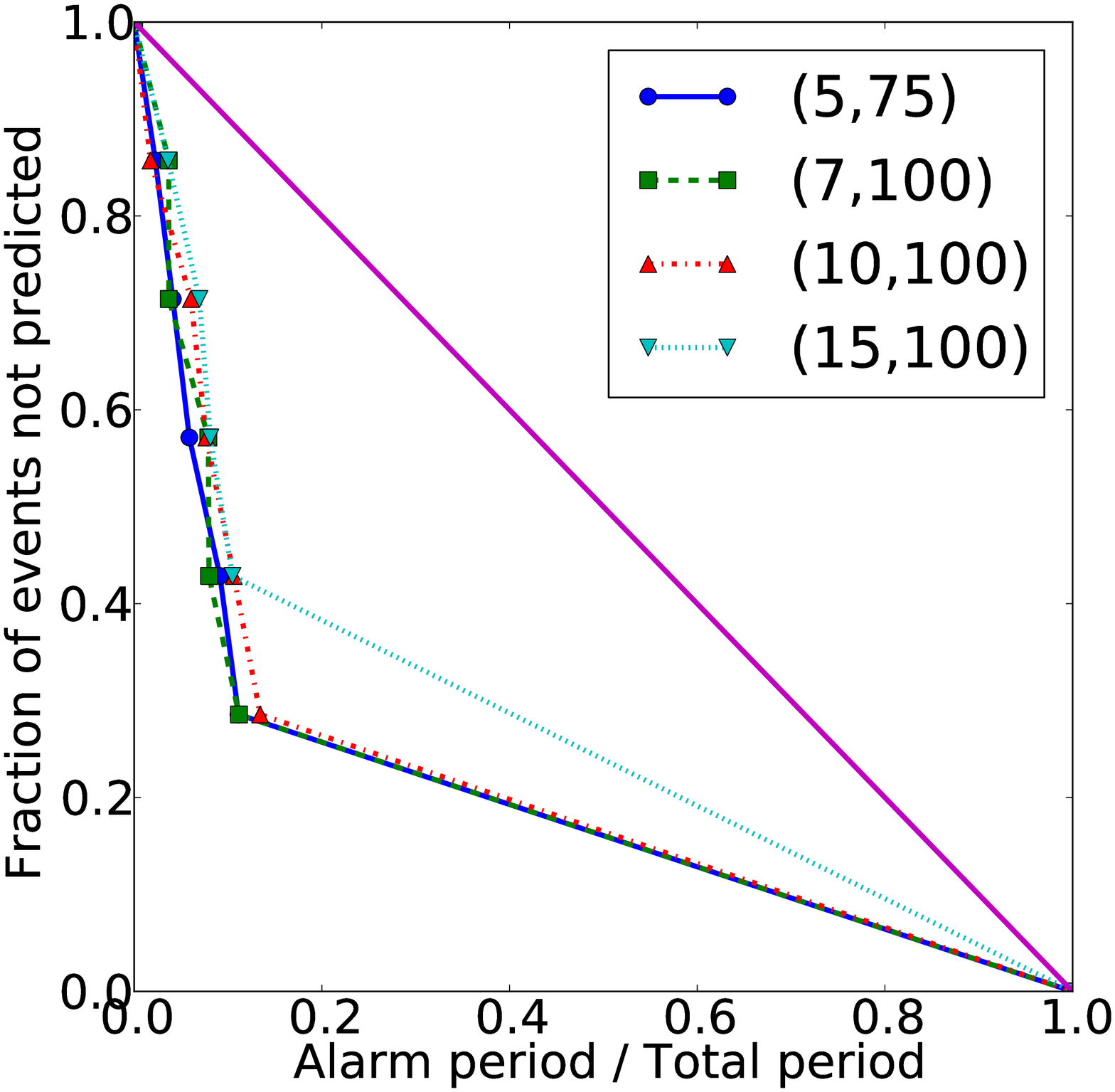}
\caption{{\bf Error diagram for back tests and predictions of crashes and
rebounds for Nikkei 225 index with different types of feature qualifications.}
The format is the same as Fig.~\ref{fig:errrussell}. } \label{fig:errnikkei}
\end{figure}

In Figs.~\ref{fig:errrussell}--\ref{fig:errnikkei}, we show the results on
predictions and back tests in terms of error diagrams for crashes and rebounds
in each of the indices. The results for different feature qualification pairs
$(\alpha, \beta)$ are shown in each figure. All these figures show that our
alarm index for crashes and rebounds in either back testing or prediction
performs much better than random.

[Figs.~\ref{fig:errrussell}--\ref{fig:errnikkei} about here.]

\section{Trading Strategy} \label{sec:strategy}

One of the most powerful methods to test the predictability of a signal is to
design simple trading strategies based on it.  We do so with our alarm index by
using simple moving average strategies, which keep all the key features of the
alarm index and avoid parametrization problems.  The strategies are kept as
simple as possible and can be applied to any indices.

The trading strategies are designed as follows: the daily exposure of our
strategy $\theta$ is determined by the average value of the alarm index for the
past $n$ days. The rest of our wealth, $1-\theta$, is invested in a 3-month US
treasury bill.

Let us denote the average rebound and crash alarm index of the past $n$ days as $AI_R$
and $AI_C$ respectively. We create three different strategies:
\begin{itemize}
\item A long strategy using only the rebound alarm index. We will take a long
position in this strategy only. The daily exposure of our strategy is based on
the average value of the past $n$ days rebound alarm index: $\theta = AI_R$.
\item A short strategy using only the crash alarm index. We will take a short
position in this strategy only. The daily absolute exposure of our strategy is
based on the average value of the past $n$ days crash alarm index: $|\theta| =
AI_C$.
\item A long-short strategy linearly combining both strategies above. When the
average rebound alarm index is higher than the average crash alarm index, we
take a long position and vice versa. The absolute exposure $|\theta| = |AI_R -
AI_C|$.
\end{itemize}

These strategies have the advantage of having few parameters, as only the
duration $n$ needs to be determined. Despite their simplicity, they capture the
two key features of the alarm index. First, we see that the alarms are
clustered around certain dates.  The more clustering seen, the more likely that
a change of regime is coming and, therefore, the more we should be invested.
Second, we see that a strong alarm close to $1$ should be treated as more
important than a weaker alarm while at the same time the smaller alarms still
contain some information and should not be discarded.

Tab.~\ref{tab:allstratsharpes} summarizes the Sharpe ratios for long-short
strategies on the out-of-sample period (testing set) for each index. The
strategy is calculated with four different moving average look-backs: $ n = 20,
30, 40, 60$ days. We use the Sharpe ratios of the market during this period as
the benchmark in this table. Recall that the Sharpe ratio is a measure of the
excess return (or risk premium) per unit of risk in an investment asset or a
trading strategy. It is defined as:
\begin{equation}
 S = \frac{R-R_f}{\sigma} = \frac{R-R_f}{\sqrt{Var[R-R_f]}}~,
\end{equation}
where $R$ is the return of the strategy and $R_f$ is the risk free rate. We use
the US three-month treasury bill rate here as the risk free rate. The Sharpe
ratio is used to characterize how well the return of an asset compensates the
investor for the risk taken: the higher the Sharpe ratio number, the better.
When comparing two assets with the same expected return against the same risk
free rate, the asset with the higher Sharpe ratio gives more return for the
same risk. Therefore, investors are often advised to pick investments with high
Sharpe ratios. From Tab.~\ref{tab:allstratsharpes}, we can find that, for seven
out of ten global major indices, the Sharpe ratios of our strategies (no matter
which look-back duration $n$ is chosen) are much higher than the market, which
means that our strategies perform better than the simple buy and hold strategy.
This result indicates that the JLS model combined with the pattern recognition
method has a statistically significant power in systematic detection of rebounds and of
crashes in financial markets.

[Tab.~\ref{tab:allstratsharpes} about here.]

\begin{figure}[!hb]
\centering
\includegraphics[width=0.9\textwidth]{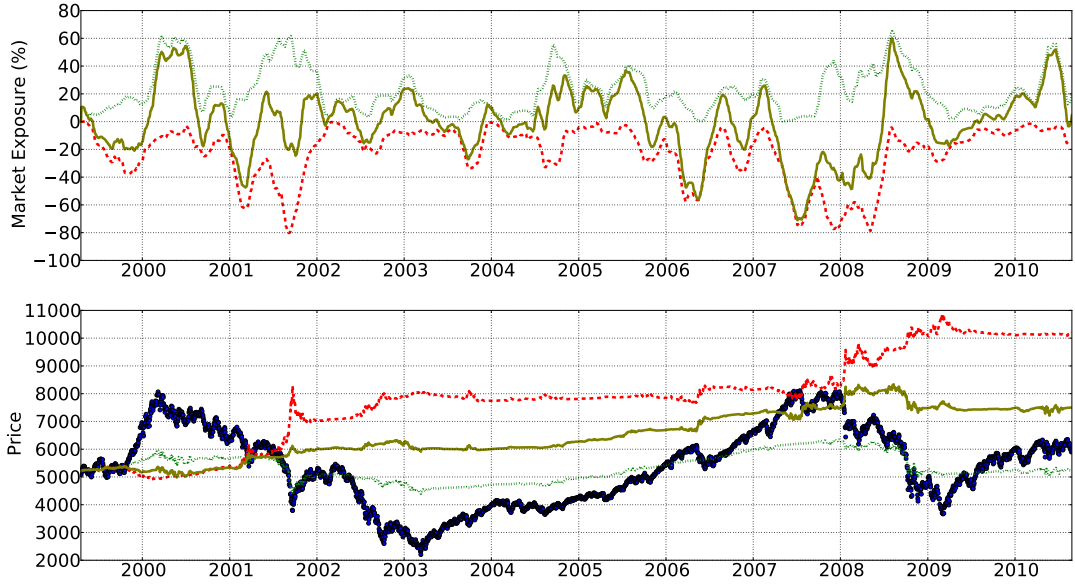}
\caption{{\bf Performance of trading strategy using our technique: DAX index 60
days moving average strategy using the feature qualification pair $(15,100)$
for crashes and $(10, 200)$ for rebounds.} (upper) the exposures for different
strategies, where the olive solid line represents the long-short strategy, the
green dotted line and red dashed line are for long and short strategy
respectively. (Lower) The historical price and wealth trajectories of the
strategies. The blue circles represent the historical price of the index while
the others are the wealth trajectories consistent with the upper figure (olive
solid - long-short, green dotted - long, red dashed - short).\\The Sharpe ratio
for the strategies are $0.07$ (long-short), $-0.66$ (long) and $0.41$ (short).
The Sharpe ratio of the corresponding benchmarks, which consist of constant
position in the market with exposure equal to the strategy over the whole
period, are $0.06$ (long-short), $-0.06$ (long) and $0.06$ (short). And the
Sharpe ratio of the index in this period is $-0.06$. Note that the short
strategy performs better than long-short or long strategies as discussed in the
text.}
\label{fig:daxstrategy}%
\end{figure}

The long-short strategies for CAC 40, DAX and ASX perform not as well as the
market. However, this is not a statement against the prediction power of
our method, but instead supports the evidence that our method detects
specific signatures preceding rebounds and crashes that are essentially
different from high volatility indicators. Contrary to common lore
and to some exceptional empirical cases \cite{schwert_stock_1990,AndersenSor04},
crashes of the financial markets often happen during
low-volatility periods and terminate them. By construction, both our rebound alarm index and the crash alarm index are high
during such volatile periods. So if the alarm indices of both types are high at
the same time, it is likely that the market is experiencing a highly volatile
period. Now, we can refine the strategy and combine the evidence of
a directional crash or rebound, together with a high volatility indicator.
If we interpret the two co-existing evidences as a signal for a crash,
we should ignore these rebound alarm index and
take the short strategy mentioned before. As an application, we show the wealth
trajectories of DAX based on different type of strategies in
Fig.~\ref{fig:daxstrategy}. In the beginning of 2008, both the rebound alarm
index and the crash alarm index for DAX are very high, therefore, we detect
this period as a highly volatile period and ignore the rebound alarm index. The
short strategy gives a very high Sharpe ratio $S = 0.41$ compared to the
long-short and short benchmarks where $S = 0.06$. The strategy's average weight
is used to compute these benchmarks. These simple benchmarks are constantly
invested by a given percentage in the market so that, over the whole time
period, they give the same exposure as the corresponding strategy being tested
but without the genuine timing information the strategy should contain.

[Figs.~\ref{fig:daxstrategy} about here.]

\begin{figure}[!hb]
\centering
\includegraphics[width=0.9\textwidth]{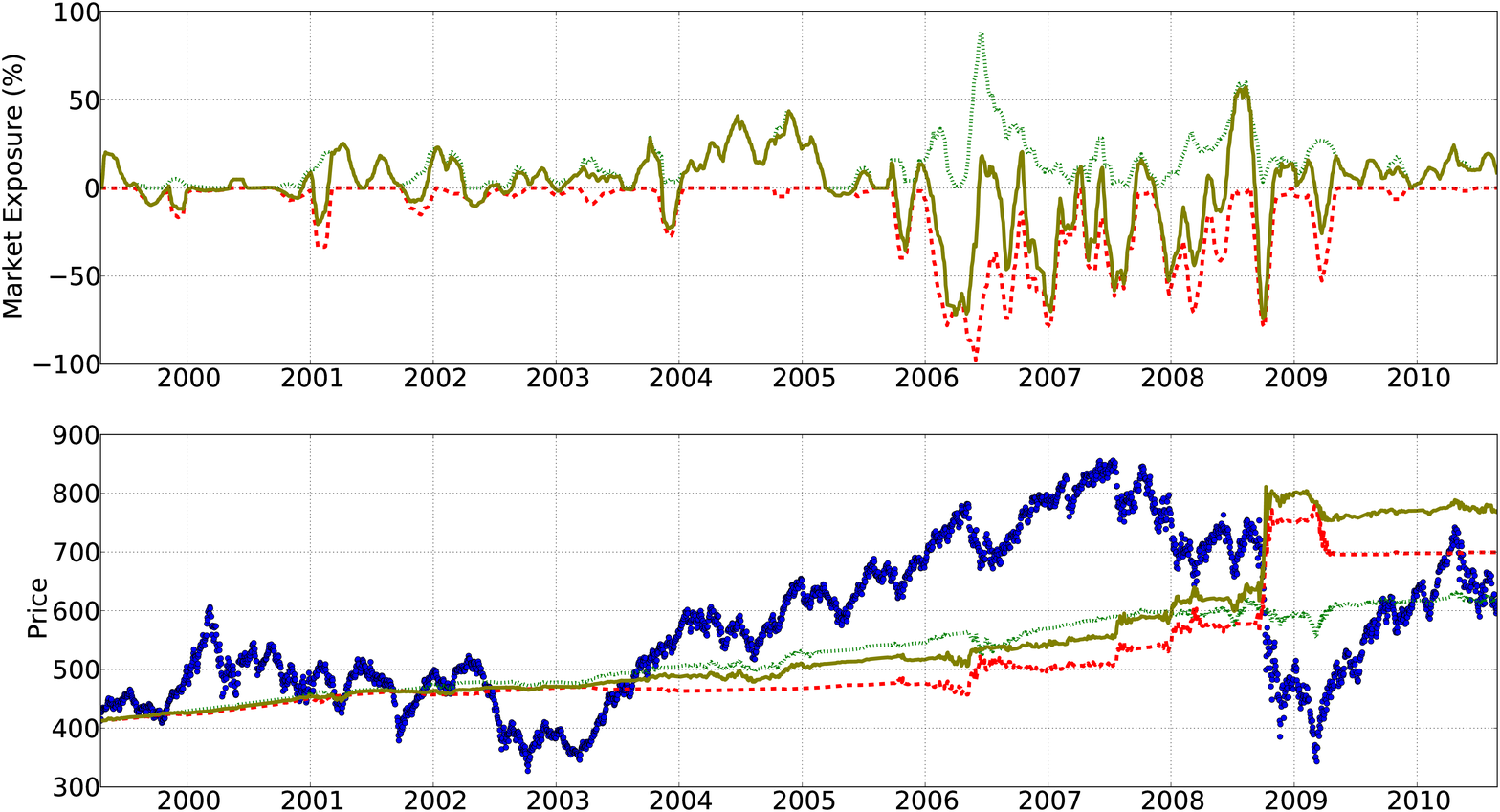}
\caption{{\bf Performance of trading strategy using our technique: Russell 2000
index 30 days moving average strategy using the feature qualification pair
$(15,100)$ for crashes and $(10, 100)$ for rebounds.} The format is the same as
Fig.~\ref{fig:daxstrategy}. \\The Sharpe ratio for the strategies are $0.47$
(long-short), $0.18$ (long) and $0.26$ (short). The Sharpe ratio of the
corresponding benchmarks, which consist of constant position in the market with
exposure equal to the strategy over the whole period, are $0.03$ (long-short),
$0.03$ (long) and $-0.03$ (short). And the Sharpe ratio of the index in this
period is $0.03$.}
\label{fig:rutstrategy}%
\end{figure}

\begin{figure}[!hb]
\centering
\includegraphics[width=0.9\textwidth]{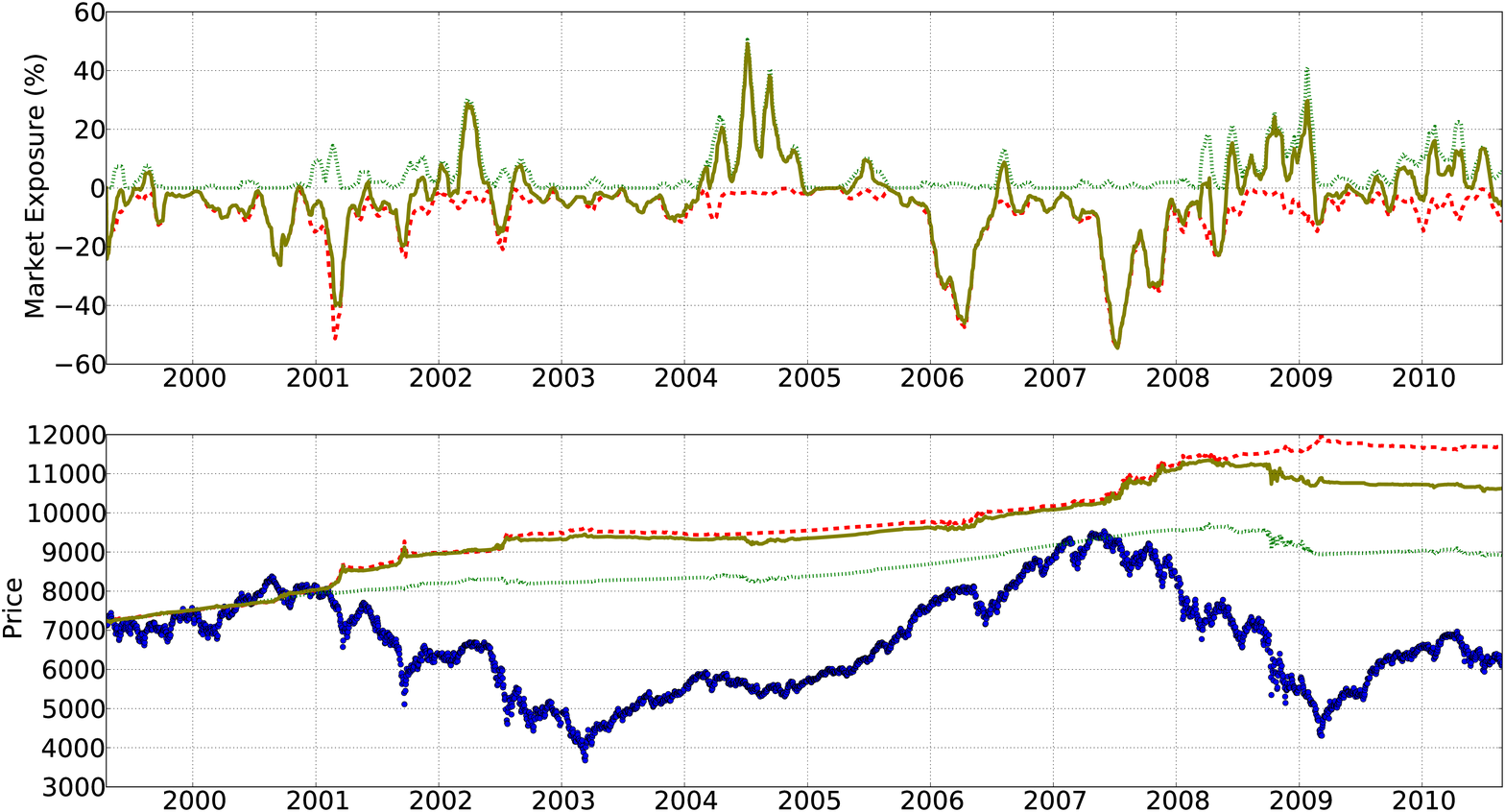}
\caption{{\bf Performance of trading strategy using our technique: SMI index 20
days moving average strategy.} The format is the same as
Fig.~\ref{fig:daxstrategy}. \\The Sharpe ratio for the strategies are $0.28$
(long-short), $-0.04$ (long) and $0.65$ (short). The Sharpe ratio of the
corresponding benchmarks, which consist of constant position in the market with
exposure equal to the strategy over the whole period, are $0.2$ (long-short),
$-0.2$ (long) and $0.2$ (short). And the Sharpe ratio of the index in this
period is $-0.2$.}

\label{fig:ssmistrategy}%
\end{figure}

\begin{figure}[!hb]
\centering
\includegraphics[width=0.9\textwidth]{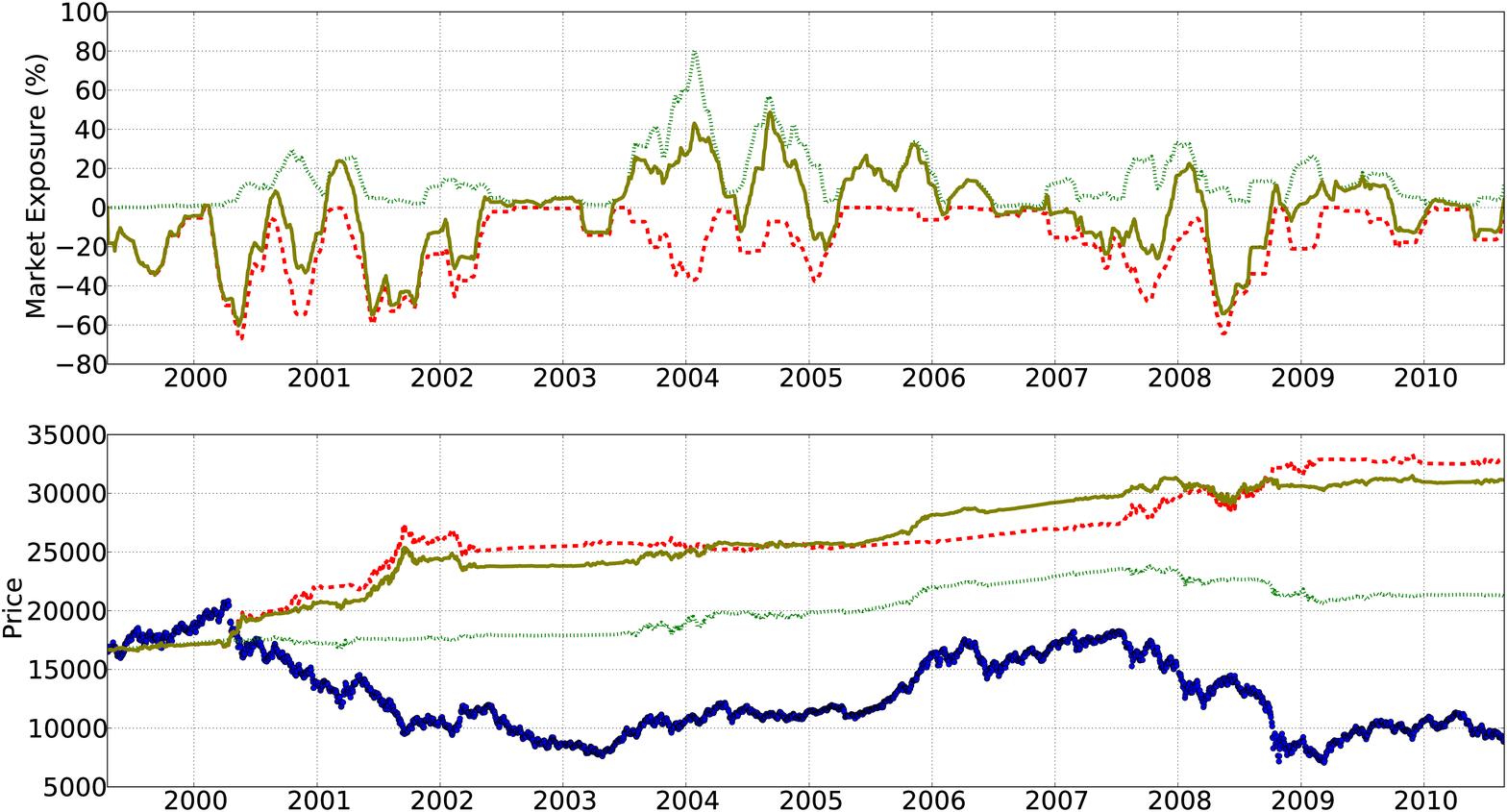}
\caption{{\bf Performance of trading strategy using our technique: Nikkei 225
index 60 days moving average strategy.} The format is the same as
Fig.~\ref{fig:daxstrategy}. \\The Sharpe ratio for the strategies are $0.59$
(long-short), $-0.11$ (long) and $0.59$ (short). The Sharpe ratio of the
corresponding benchmarks, which consist of constant position in the market with
exposure equal to the strategy over the whole period, are $0.33$ (long-short),
$-0.33$ (long) and $0.33$ (short). And the Sharpe ratio of the index in this
period is $-0.33$.}

\label{fig:n225strategy}%
\end{figure}

As before, the detailed out-of-sample performances for each sample index
(Russell 2000, SMI and Nikkei 225) are also tested.
Fig.~\ref{fig:rutstrategy}--\ref{fig:n225strategy} illustrate the wealth
trajectories for different strategies. In order to show the consistency of the
strategies with respect to the chosen parameters in the pattern recognition
method, we show the performance of the Russell 2000 index for different
qualification pairs: $(15, 100)$ for rebounds and $(10,100)$ for crashes. From
these wealth trajectories, it is very obvious that our alarm indices can catch
the market rebounds and crashes efficiently.

[Figs.~\ref{fig:rutstrategy}--\ref{fig:n225strategy} about here.]

The detailed performances for these three stock indices are listed in
Tabs.~\ref{tab:rutperf1}--\ref{tab:n225perf}. We also provide the performance
of the Russell 2000 index for the `normal' qualification pair: $(7, 100)$ in
Tab.~\ref{tab:rutperf2} as a reference. These tables confirm again that
strategies mostly succeed in capturing big changes of regime. Compared to the
market, the strategies based on our alarm index perform better than the market
for more than eleven years in all the important measures: Annual returns and
Sharpe ratios are larger, while volatilities, downside deviations and maximum
drawdowns are smaller than the market performance.

[Tabs.~\ref{tab:rutperf1}--\ref{tab:n225perf} about here.]

\section{Conclusion}\label{sec:conclusion}

We provided a systematic method to detect financial crashes and rebounds. The
method is a combination of the JLS model for bubbles and negative bubbles, and
the pattern recognition technique originally developed for earthquake
predictions. The outcome of this method is a rebound/crash alarm index to
indicate the probability of a rebound/crash for a certain time. The
predictability of the alarm index has been tested by 10 major global stock
indices. The performance is checked quantitatively by error diagrams and
trading strategies. All the results from error diagrams indicate that our
method in detecting crashes and rebounds performs better than chance and
confirm that the new method is very powerful and robust in the prediction of
crashes and rebounds in financial markets. Our long-short trading strategies
based on the crash and rebound alarm index perform better than the benchmarks
(buy and hold strategy with the same exposure as the average exposure of our
strategies) in seven out of ten indices. Highly volatile periods are observed
in the indices of which the long-short trading strategy fails to surpass the
benchmark. By construction of the alarm index and the fact that highly volatile
periods are not coherent with bullish markets, we claim that we should
ignore the rebound alarm index during such volatile periods. This statement has
been proved by the short strategy which only consider the crash alarm index.
Thus, our trading strategies confirm again that the alarm index has a strong
ability in detecting rebounds and crashes in the financial markets.

\clearpage
\begin{table}[h]
  \centering
    \begin{tabular}{llrrrr}
    \toprule[2pt]
    Index name  & Yahoo ticker & Learning start   & Prediction start    & Prediction end  & win. \# \\
    \midrule[1pt]

    S\&P 500 & \^{}GSPC & 5-Jan-1950 & 26-Mar-1999 & 3-Jun-2009 & 11662 \\
    Nasdaq & \^{}IXIC & 13-Dec-1971 & 20-Mar-1999 & 30-Jul-2010 & 7209 \\
    Russell 2000 & \^{}RUT & 30-Sep-1987 & 17-Apr-1999 & 27-Aug-2010 & 4270 \\
    FTSE 100 & \^{}FTSE & 3-May-1984 & 17-Apr-1999 & 27-Aug-2010 & 4970 \\
    CAC 40 & \^{}FCHI & 1-Mar-1990 & 17-Apr-1999 & 27-Aug-2010 & 3766 \\
    SMI   & \^{}SSMI & 9-Nov-1990 & 17-Apr-1999 & 27-Aug-2010 & 3626 \\
    DAX   & \^{}GDAXI   & 26-Nov-1990 & 17-Apr-1999 & 27-Aug-2010 & 3626 \\
    Nikkei 225 & \^{}N225 & 4-Jan-1984 & 17-Apr-1999 & 27-Aug-2010 & 5026 \\
    Hang Seng & \^{}HSI & 31-Dec-1986 & 17-Apr-1999 & 27-Aug-2010 & 4410 \\
    ASX       & \^{}AORD    & 6-Aug-1984  & 17-Apr-1999 & 27-Aug-2010 & 4914 \\
    \bottomrule[2pt]
    \end{tabular}%
    \caption{{\bf Information for the tested indices}: Yahoo ticker of each index,
     starting time of learning and prediction periods, ending time of prediction
      and number of sub-windows.}
  \label{tab:numintervals}%
\end{table}%

\begin{table}[h]
  \centering
    \begin{tabular}{lrrrrr}
    \toprule[2pt]
    Index   & \multicolumn{4}{c}{ Strategy (duration $n$)} & Market \\
    \cline{3-6}
           & 20 & 30 & 40 & 60 &\\
    \midrule[1pt]
    S\&P 500     & 0.32  & 0.38  & 0.43  & 0.39  & -0.28 \\
    Nasdaq      & 0.18  & 0.41  & 0.48  & 0.34  & -0.11 \\
    Russell 2000  & 0.29  & 0.3   & 0.27  & 0.27 & 0.03\\
    FTSE 100     & -0.07 & -0.06 & 0.01  & 0.05 & -0.22\\
    CAC 40       & -0.18 & -0.35 & -0.37 & -0.24& -0.19\\
    SMI            & 0.28  & 0.22  & 0.14  & 0.13 &  -0.20\\
    DAX           & -0.26 & -0.24 & -0.27 & -0.19& -0.06\\
    Nikkei 225    & 0.07  & 0.19  & 0.39  & 0.59 & -0.33\\
    Hang Seng      & 0.32  & 0.38  & 0.31  & 0.23 & 0.06\\
    ASX          & -0.38 & -0.41 & -0.33 & -0.24& 0.02\\
    \bottomrule[2pt]
    \end{tabular}%
  \caption{{\bf Summary of Sharpe ratios for the market and
  the long-short strategies with different moving average duration $n$.}
  The start and end dates of
  the strategies are 26-Mar-1999 -- 3-Jun-2009 for S\&P 500, 20-Mar-1999 -- 30-Jul-2010
  for Nasdaq and 17-Apr-1999 -- 27-Aug-2010 for others.
  The feature qualification pairs of $(7,100)$ for both crash and rebound alarm indices are
used in this calculation.}
  \label{tab:allstratsharpes}%
\end{table}%

\begin{table}[!ht]
  \begin{center}
    \begin{tabular}{lrrrrr}
    \toprule[2pt]
     & \multicolumn{4}{c}{Strategy (duration $n$)}& Market\\
    \cline{2-5}
    & 20 & 30 & 40 & 60& \\
    \midrule[1pt]
    Ann Ret  & 5.3\% & 5.4\% & 4.8\% & 4.2\%&0.1\% \\
    Vol   & 6.4\% & 5.9\% & 5.3\% & 4.1\%  &26.3\%\\
    Downside dev &  4.2\% & 3.6\% & 3.4\% & 2.7\%&18.9\% \\
    Sharpe & 0.42  & 0.47  & 0.41  & 0.37 &0.03\\
    Max DD &10\%  & 7\%   & 7\%   & 6\% & 65\%\\
    Abs Expo & 16\%  & 15\%  & 14\%  & 12\% &  \\
    Ann turnover & 328\% & 220\% & 167\% & 123\%&  \\
    \bottomrule[2pt]
    \end{tabular}%
  \caption{{\bf Russell 2000 index long-short strategies
   out-of-sample performance table.} Start date: 17-Apr-1999,
   end date 27-Aug-2010. Qualification pairs: $(15, 100)$ for
   rebounds and $(10,100)$ for crashes.}
  \label{tab:rutperf1}%
  \end{center}
\end{table}%

\begin{table}[!ht]
  \begin{center}
    \begin{tabular}{lrrrrr}
    \toprule[2pt]
     & \multicolumn{4}{c}{Strategy (duration $n$)}& Market\\
    \cline{2-5}
     & 20 & 30 & 40 & 60& \\
    \midrule[1pt]
    Ann Ret  & 4.8\% & 4.7\% & 4.3\% & 4.0\% &0.1\%\\
    Vol    & 7.9\% & 7.1\% & 6.5\% & 5.2\% &26.3\%\\
    Downside dev & 5.4\% & 4.8\% & 4.5\% & 3.6\% &18.9\% \\
    Sharpe & 0.29  & 0.30  & 0.27  & 0.27&0.03 \\
    Max DD &11\%  & 12\%  & 9\%   & 9\% & 65\%\\
    Abs Expo & 23\%  & 21\%  & 20\%  & 18\% &  \\
    Ann turnover & 446\% & 306\% & 237\% & 172\%&  \\
    \bottomrule[2pt]
    \end{tabular}%
    \caption{{\bf Russell 2000 index long-short strategies
    out-of-sample performance table.} Start date: 17-Apr-1999,
    end date 27-Aug-2010. Qualification pairs: $(7, 100)$ for
    both rebounds and crashes.}
   \label{tab:rutperf2}%
  \end{center}
\end{table}%

\begin{table}[!ht]
  \begin{center}
    \begin{tabular}{lrrrrr}
    \toprule[2pt]
    & \multicolumn{4}{c}{Strategy (duration $n$)}& Market \\
    \cline{2-5}
     & 20 & 30 & 40 & 60 &\\
    \midrule[1pt]
    Ann Ret & 3.4\% & 3.3\% & 3.0\% & 3.0\% &-3.4\%\\
    Vol  & 2.6\% & 2.6\% & 2.5\% & 2.3\%  &20.4\% \\
    Downside dev & 1.7\% & 1.6\% & 1.6\% & 1.5\%&14.6\%\\
    Sharpe & 0.28  & 0.22  & 0.14  & 0.13 & -0.20\\
    Max DD &7\%   & 6\%   & 7\%   & 5\% & 59\%  \\
    Abs Expo & 9\%   & 9\%   & 9\%   & 9\%  &  \\
    Ann turnover & 134\% & 102\% & 77\%  & 55\%& \\
    \bottomrule[2pt]
    \end{tabular}%
    \caption{{\bf SMI index long-short strategies out-of-sample performance
    table.} Start date 17-Apr-1999, end date 27-Aug-2010.}
   \label{tab:ssmiperf}%
  \end{center}
\end{table}%

\begin{table}[!ht]
  \begin{center}
    \begin{tabular}{lrrrrr}
    \toprule[2pt]
    & \multicolumn{4}{c}{Strategy (duration $n$)} & Market \\
    \cline{2-5}
    & 20 & 30 & 40 & 60 &\\
    \midrule[1pt]
    Ann Ret & 3.0\% & 3.7\% & 4.8\% & 5.7\% &-8.4\% \\
    Vol   & 6.3\% & 6.0\% & 5.6\% & 5.0\% & 25.4\%\\
    Downside dev & 4.4\% & 4.1\% & 3.7\% & 3.2\%&18.7\% \\
    Sharpe &0.07  & 0.19  & 0.39  & 0.59  & -0.33 \\
    Max DD & 15\%  & 13\%  & 10\%  & 8\% & 75\%   \\
    Abs Expo & 19\%  & 18\%  & 17\%  & 16\%& \\
    Ann turnover & 417\% & 300\% & 223\% & 146\%& \\
    \bottomrule[2pt]
    \end{tabular}%
    \caption{{\bf Nikkei 225 index long-short strategies out-of-sample performance
    table.} Start date 17-Apr-1999, end date 27-Aug-2010.}
   \label{tab:n225perf}%
  \end{center}
\end{table}%

\clearpage

\clearpage

\clearpage

\clearpage

\end{document}